\title{Consistent Calibration of Economic Scenario Generators: the Case for Conditional Simulation}
\author{
        Misha van Beek
}
\date{\today}

\documentclass[12pt]{article}
\usepackage[margin=1in]{geometry}

\usepackage{url}

\usepackage{breakurl}
\usepackage[breaklinks]{hyperref}

\usepackage{amsfonts}
\usepackage{amsmath}
\allowdisplaybreaks

\usepackage{tikz}
\usetikzlibrary{calc}
\usetikzlibrary{shapes.multipart}
\usetikzlibrary{arrows,decorations.markings,positioning,arrows.meta} 

\usepackage[longend,ruled]{algorithm2e}

\usepackage{enumitem}

\usepackage{booktabs,array,adjustbox,dcolumn}
\newcolumntype{A}{ >{$} r <{$} @{} >{${}} l <{$} } 
\newcolumntype{R}[2]{>{\adjustbox{angle=#1,lap=\width-(#2)}\bgroup}l<{\egroup}}
\newcolumntype{C}[1]{>{\centering\let\newline\\\arraybackslash\hspace{0pt}}m{#1}}

\usepackage[font={footnotesize}]{caption}

\usepackage{natbib}
\DeclareMathOperator{\vcov}{vcov}
\DeclareMathOperator{\E}{\mathbb{E}}

\begin{document}
\maketitle

\begin{abstract}
\noindent Economic Scenario Generators (ESGs) simulate economic and financial variables forward in time for risk management and asset allocation purposes. It is often not feasible to calibrate the dynamics of all variables within the ESG to historical data alone. Calibration to forward-information such as future scenarios and return expectations is needed for stress testing and portfolio optimization, but no generally accepted methodology is available. This paper introduces the Conditional Scenario Simulator, which is a framework for consistently calibrating simulations and projections of economic and financial variables both to historical data and forward-looking information. The framework can be viewed as a multi-period, multi-factor generalization of the Black-Litterman model, and can embed a wide array of financial and macroeconomic models. Two practical examples demonstrate this in a frequentist and Bayesian setting.
\end{abstract}

\renewcommand{\abstractname}{Acknowledgements}
\begin{abstract}
\noindent I would like to thank Andrew Ang, Jean Boivin, Linxi Chen, Bingxu Chen, David Greenberg, Michel Mandjes, Peter Spreij and Erik Winands for their great help and suggestions on previous drafts of this paper.
\end{abstract}

\section{Introduction}


Economic Scenario Generators (ESGs) are models that simulate economic and financial variables forward in time. They are primarily used to analyse existing asset allocations and balance sheets of financial institutions such as banks, insurers and asset managers against stressed scenarios. Financial institutions are often required to do so by regulators. A second application is in the construction of new allocations. ESGs can simulate the movements of financial markets that feed into the portfolio optimization process. 

ESGs typically consist of many sub-models that all have parameters to be set. These parameters are usually calibrated to historical data. Historical data alone is insufficient for two applications. First, in stress testing regulators prescribe partial calibrations in the form of forward-looking information about the economy. These calibrations are far from historical averages by design, and the onus is on the user to \emph{consistently calibrate} all other quantities. Second, when using an ESG for portfolio optimization, it needs to be calibrated against (often expert-based) views known as Capital Market Assumptions (CMAs). These are views on the mean returns of primary asset classes. If all other asset mean return calibrations are not consistent with these CMAs, then optimization routines will return highly concentrated allocations. For example, if two strongly correlated equities have diverging mean return calibrations, then an extreme long-short position can theoretically (but rarely in practice) achieve high returns with low volatility.

There is no generally accepted approach to consistently calibrate ESGs to historical data and forward-looking information simultaneously. To address this gap, this paper introduces the Conditional Scenario Generator (CSG) as a framework for prediction, stress testing and asset allocation. Similar to an ESG, the CSG allows for joint analysis of macroeconomic variables, financial factors and asset expected and realized returns in a multi-period context, where forecasts are driven by dynamics fitted on historical data. But the CSG \emph{embeds a structured approach to calibration} to forward-looking information such as stressed scenarios or CMAs expressed as expert views.

For a practical example of the role of the CSG in an investment process, consider the following case. Suppose an investor has to make a strategic asset allocation decision across several portfolios and wants to know their mean returns to this end. The investor has several medium-term views on macroeconomic variables such as GDP growth and future policy rates, as well as CMAs in the form of long-term views on the mean returns of major asset classes. The CSG can determine what the mean returns are on each portfolio conditional on all views at each horizon.

Next, suppose the investor is worried about a stressed scenario in which a demand-driven recession hits the economy. Such a scenario can be expressed as a negative economic growth shock, in combinations with low inflation. The CSG can be used to simulate price paths that are consistent with a specific set of assumptions, e.g.~-2\% quarterly GDP growth and 0\% consumer price index growth at a 2 year horizon. This shows whether the chosen asset allocation is robust to such a scenario.

Asset allocation problems conditional on views of mean returns are often solved using the Black-Litterman (BL) model \citep{black1992}. However, this allows the investor only to express views on mean returns of assets at one prespecified horizon, as (A) macroeconomic variables are not integrated into the model, the model is (B) underpinned by a single-factor explanation of the market, and (C) single-period in nature. In contrast, as a generalization of the BL model to a multi-period, multi-factor and macro-informed framework, the CSG can synthesize more diverse information into the mean return predictions, and derive term structures of return expectations rather than point forecasts.

As mentioned above, stress testing an existing allocation or portfolio is a form of scenario analysis that is at the core of modern regulation such as ORSA, CCAR, DFAST, CECL in the US, and Solvency, Basel and IFRS9 in Europe \citep{acharya2012capital,cole2014basis}. These tests require institutions to project losses given macroeconomic scenarios that the regulator explicitly provides, or require institutions to come up with their own scenarios tailored to their portfolios. Since the CSG jointly models macroeconomic and financial variables, it is relatively straightforward to calibrate to macroeconomic variables to see portfolio losses, or (in the so-called reverse stress test), condition on portfolio losses to see what macroeconomic environment explains them best \citep{grundke2011reverse,breuer2012systematic}. 

The mechanics of the CSG are based on analytical (Kalman) and simulation smoothing in a dynamic linear model (DLM).\footnote{Also often referred to as a state-space model, although the terminology is somewhat fuzzy.} The DLM setup incorporates both a macroeconomic model and a financial markets model, that are tied together with a linear macro-financial link. Most popular macroeconomic models, such as the vector auto-regressive (VAR) family and (log-linearized) dynamic stochastic general equilibrium (DSGE) models can be written in the form of a DLM when joint normality is assumed. The financial markets model follows the classical setup of a linear factor model to explain asset returns. In this way, the CSG encapsulates the BL model as a special case, with the same predicted mean returns for specific settings that are explained in Appendix \ref{sec:blacklitterman}.

Despite regulatory emphasis, the existing literature on generating calibrated scenarios is thin at best. \citet{golub2018market} point out that no generally accepted framework exists, and that the research on best practises is limited \citep{clemen1999combining}.

\citet{golub2018market} propose a framework for calibrating asset returns to financial scenarios. Their Market-Driven Scenario (MDS) approach follows the conditioning philosophy outlined by \citet{kupiec2002stress}. The core concept is to consider the joint distribution of factors that drive financial outcomes, and look at the conditional distribution of outcomes given an explicit value for a subset of these factors that capture the scenario. This is a powerful and practical idea, but it is not directly applicable to answer regulatory questions. First, it is unclear how to extend the regulatory scenarios that are described in mostly macroeconomic terms to financial factors. Second, regulatory scenarios are multi-period and cannot easily be flattened into a single-period equivalent. The CSG can be viewed as a multi-period extension of the MDS approach to macroeconomic quantities.

For portfolio construction, the BL model is a close cousin of the MDS approach. But other extensions to the BL model exist that allow for calibration of future financial outcomes against expert views. \citet{meucci2010} notes that these views can also represent scenarios for stress testing purposes. Most of these extensions focus on generalizing the distributional assumptions underpinning the model, as well as the financial quantities that the user can have views on. For example, through a modification of existing simulations called entropy pooling, \citet{meucci2008fully} shows how to obtain a sample from a posterior distribution given highly general non-linear views that can be expressed on volatilities and correlations as well as macroeconomic quantities. These views may also apply to financial factors \citep{meucci2009}. The extension that the entropy pooling technique gives is clearly beneficial in terms of the flexibility of the views that can be incorporated. While this is important, it is still a single-period framework that cannot handle the multi-period nature of macroeconomic scenarios. Related approaches that are not discussed here share this shortcoming \citep{qian2001conditional, pezier2007global,almgren2007optimal,palczewski2019black}. In contrast, the CSG is multi-period in nature, but does not address the non-normality and non-linearity of certain views. The CSG is thus more limited in the breadth of views themselves that it can express, but less limited in their timing.

Outside the portfolio and risk management context, calibrating a model to multi-period scenarios is more common. Macroeconomists are usually interested in gauging the impact of a government policy or of a macroeconomic shock on the economy or financial variables of interest. To this end they calibrate models to an impulse, or more generally to a set of shocks, to obtain impulse response functions. A standard way to do this is through analytical (Kalman) smoothing of a DLM. The smoother computes the marginal distribution of variables at each horizon, conditional on all past, present and future information, in a jointly normal setup that works for a wide range of macroeconomic models \citep[see][for technical details]{clarida1984conditional,waggoner1999conditional,banbura2015conditional}. There are myriad examples of analyses that use this approach \citep[for example][]{jarocinski2008house,giannone2008business,lenza2010monetary,bloor2011real,giannone2012ecb,giannone2014short}. The CSG follows a similar smoothing approach, but models the behavior of assets explicitly. Even when financial variables are included, such as by \citet{ha2020global}, there is no specific model of the financial markets available to simulate financial outcomes that are directly relevant to asset managers. The CSG includes a model of financial markets that is linked to the economy, such that assets can be priced consistently in this framework.

To the best of my knowledge, there exists one other framework that allows consistent calibration of multi-period, macro-consistent simulations and that also contains the appropriate structure to model relevant financial outcomes. \citet{vanderschans2017} propose a time-dependent generalization of the Black-Litterman framework, which includes a multi-factor model. In their definition of what a factor is, they include macroeconomic variables. This model is different primarily in three shortcomings that the CSG addresses. First, \citeauthor{vanderschans2017} require a specific statistical factor model that merges both macroeconomic and financial variables. The power of the CSG is that it can build on existing macroeconomic and factor models from a broad class. Second, there is no distinction between financial factors and assets. These are mixed, which means that implicitly the exposures of assets are determined through regression. Unlike in the CSG framework, assets with time-varying exposures to underlying risk factors, such as bonds, cannot be included in the analysis. Third, forecasts are not impacted by views at later horizons, and hence their framework is not fully forward-looking. For example a high-rates view at time 5 would see business-as-usual forecast at time 4, with a sudden jump to time 5. This is unrealistic as rates tend to hike, not jump, and is problematic in particular for scenario analysis.

This paper is organized as follows. Section \ref{sec:notation} outlines the structure of the model, introduces the macroeconomic and financial market components, and explains the link between these two. This section concludes with the conditioning framework. Section \ref{sec:estimation} discusses how the different components of the model can be estimated on historical data, and how an externally estimated model can be brought into the analysis under certain assumptions. Section \ref{sec:egB} and \ref{sec:egC} discuss fully estimated examples of the framework; the first from a frequentist and the second from a Bayesian perspective. I conclude in Section \ref{sec:conclusion}. The connection with the Black-Litterman model, as well as the mathematics behind the conditional forecasting algorithms are available in the Appendix.

\section{Notation of the general framework}\label{sec:notation}

This section derives the CSG as a general calibration framework for prediction and scenario analysis in financial markets. The CSG consists of roughly three components, or models, depicted in Figure \ref{fig:framework}. The first component is a \emph{macroeconomic model} that describes the economy. The second component is a \emph{factor model}, and the third component is an \emph{asset model}, linked by exposures of the assets to the factors. Jointly, the factor and asset model are the \emph{financial markets model}. A \emph{macro-financial linkage} between the macroeconomic model and the financial markets model describes how the two domains interact. This section describes each component in detail. Jointly, these components lock down the dynamics of all hidden and observable time-series. 

The bottom block in Figure \ref{fig:framework} depicts \emph{conditioning}. Conditioning is how we can calibrate the dynamics to views on any variable within the framework to update the forecasts of these time-series with scenarios or CMAs. The support for conditioning is what brings out the power of the framework for scenario analysis and incorporating investor views, and is described at the end of this section.

\tikzstyle{abstract}=[rectangle, draw=black, text centered, anchor=north, text width=3.7cm,inner sep=1.5ex]
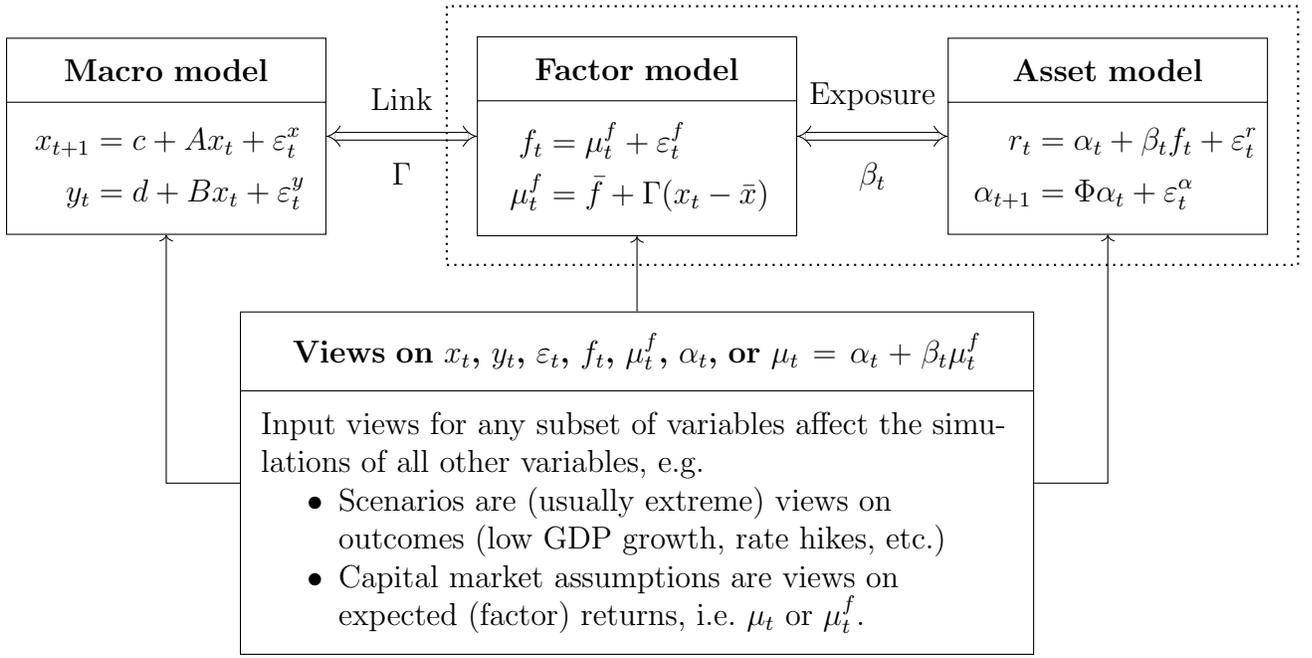
\begin{figure}[t]
\centering
\begin{tikzpicture}[node distance = 1cm and 2cm, auto]

       \node (macro) [abstract, rectangle split, rectangle split parts=2, minimum height=6cm]
        {
            \textbf{Macro model}
            \nodepart{second}
            $\begin{aligned}
            x_{t+1}&=c+Ax_t+\varepsilon_t^x\\\phantom{\mu_t^f}y_t&=d+Bx_t+\varepsilon_t^y
            \end{aligned}$
            
        };
        \node (factor) [abstract, rectangle split, rectangle split parts=2, right=of macro]
        {
            \textbf{Factor model}
            \nodepart{second}
            $\begin{aligned}
                f_t&=\mu_t^f+\varepsilon_t^f\\
                \mu_t^f&=\bar{f}+\Gamma(x_t-\bar{x})
            \end{aligned}$
        };
        \node (asset) [abstract, rectangle split, rectangle split parts=2, right=of factor]
        {
            \textbf{Asset model}
            \nodepart{second}
            $\begin{aligned}
             r_t&=\alpha_t+\beta_tf_t+\varepsilon_t^r\\
             \alpha_{t+1}&=\Phi\alpha_t+\varepsilon_t^\alpha\phantom{\mu_t^f}
            \end{aligned}$
        };
        \node (views) [abstract, rectangle split, rectangle split parts=2, below=of factor, text width=10cm]
        {
            \textbf{Views on $x_t$, $y_t$, $\varepsilon_t$, $f_t$, $\mu_t^f$, $\alpha_t$, or $\mu_t=\alpha_t+\beta_t\mu_t^f$}
            \nodepart[align=left]{second}
            Input views for any subset of variables affect the simulations of all other variables, e.g.
            \begin{itemize}[itemsep=-5pt,partopsep=0pt,topsep=0pt]
                \item Scenarios are (usually extreme) views on outcomes (low GDP growth, rate hikes, etc.)
                \item Capital market assumptions are views on expected (factor) returns, i.e.~$\mu_t$ or $\mu_t^f$.
            \end{itemize}
        };
        \draw[double distance=2.5pt, {Implies}-{Implies}] (macro.east) -- node [above = 0.5em] {Link} node [below = 0.5em] {$\Gamma$} (factor.west);
        \draw[double distance=2.5pt, {Implies}-{Implies}] (factor.east) -- node [above = 0.5em] {Exposure} node [below = 0.5em] {$\beta_t$} (asset.west);
        \draw[->] (views.west) -| (macro.south);
        \draw[->] (views.east) -| (asset.south);
        \draw[->] (views.north) -- (factor.south);
         \draw[thick,dotted] ($(factor.north west)+(-0.4,0.4)$)  rectangle node [above=4.7em] {\textbf{Financial markets model}} ($(asset.south east)+(0.4,-0.4)$);
\end{tikzpicture}
\caption{Graphical representation of the framework.}
\label{fig:framework}
\end{figure}

The structure of the framework can be seen as a generalization of the BL model. It extends BL in three dimensions, i.e.~(A) it is multi-period in nature, (B) it is multi-factor rather than CAPM based, and (C) it is macro-informed by incorporating a macroeconomic model. Appendix \ref{sec:blacklitterman} shows how the BL model is a special case of the CSG for specific factor and macroeconomic model choices, and a single time-period.

\subsection{The macroeconomic model}\label{sub:macro}

The first component, the macroeconomic model, assumes the following DLM format,
\begin{align}
\tilde{x}_{t+1}&=A\tilde{x}_t+G\varepsilon'_t,&\tilde{x}_1&\sim\mathcal{N}(\tilde{x}_{1|0},P_{1|0}), \label{eqn:dlm1}\\
\tilde{y}_t&=B\tilde{x}_t+H\varepsilon'_t,&\varepsilon'_t&\sim\mathcal{N}(0,I),\label{eqn:dlm2}
\end{align}
for $t=1,\ldots,T$ with present time $T$. Equation (\ref{eqn:dlm1}) is called the state equation and describe the auto-regressive dynamics of the latent macroeconomic states, that are not necessarily observable. The $n_x$-vector $x_t$ contains these latent macroeconomic states. For this and other variables the tilde denotes that the variables are measured in excess of their steady states $\bar{x}$, such that $\tilde{x}_t=x_t-\bar{x}$. Equation (\ref{eqn:dlm2}) is the measurement equation and shows how the latent macroeconomic states are observable through the $n_y$-vector $y_t$ of observable time-series. The error vectors $\varepsilon'_t$ are i.i.d.~standard multivariate Gaussian across time and describes both the measurement errors and structural shocks to the states. I label $\varepsilon_t^x=G\varepsilon_t'$ and $\varepsilon_t^y=H\varepsilon_t'$ as the structural shocks and measurement errors respectively.\footnote{For some applications, special care should be given to the construction of $G$ and $H$. There are infinite possible choices of $G$ and $H$ that lead to the same macroeconomic dynamics, but different impulse response functions on applying macroeconomic shocks. For standard VAR models ($H=O$, $B=I$), this is very easy to see. We only have data on $\vcov(\varepsilon_t^x)=GG^\top$, which has $n_{\varepsilon'}(n_{\varepsilon'}+1)/2$ elements, whereas $n_{\varepsilon'}^2$ elements of $G$ need to be identified. This is a hard but well-studied identification problem. Additional constraints can either be added recursively using Cholesky decomposition \citep{sims1980macroeconomics, christiano1999monetary}, via long-run assumption \citep{blanchard1989dynamic, fisher2006dynamic}, or via sign restrictions \citep{uhlig2005effects, arias2014inference}. While computing impulse response functions is not the purpose of this paper, impulse response functions can be seen as a special case of conditional forecasting. Section \ref{sub:conditional} details when and how this identification problem appears.} Using the assumption that $\tilde{x}_0$ is unconditionally Gaussian, the joint distribution of all variables $x_t$ and $y_t$ are Gaussian. This facilitates the notation for the conditional mean and covariance matrices
\begin{align}
x_t|y_1,\ldots,y_s&\sim\mathcal{N}(x_{t|s},P_{t|s}),&x_{t|s}&=\E[x_t|y_1,\ldots,y_s],&P_{t|s}&=\vcov[x_t|y_1,\ldots,y_s],
\end{align}
and similarly for the states in excess of their steady states, $\tilde{x}_t$.\footnote{The corner case $\tilde{x}_{1|0}$, $P_{1|0}$ represents the distribution of $\tilde{x}_1$ without conditioning on any measurements, and can be seen as a prior from a Bayesian perspective. In most practial applications, it is intuitive that $x_1$ start in its unconditional distribution, so $\tilde{x}_{1|0}=0$ and $P_{1|0}$ solves the discrete Lyapunov equation $AP_{1|0}A^\top-P_{1|0}+GG^\top=O$.}

This DLM format (\ref{eqn:dlm1}-\ref{eqn:dlm2}) may seem restrictive, but is in fact very general and includes a wide range of macroeconomic models that commonly are driven by Gaussian errors. 
Vector auto-regressions (VARs) of any order and with intercepts, structural VARs (SVARs) in reduced form, factor augmented VARs (FAVARs), and dynamic stochastic general equilibrium (DSGE) models in log-linearized format all qualify. Sections \ref{sec:egB} and \ref{sec:egC} give examples for a FAVAR model \citep[by][]{bernanke2005measuring} and a DSGE model \citep[by][]{ireland2011new}.


\subsection{The financial markets model}\label{sub:markets}

Following standard linear factor model literature, I assume mean asset returns can be explained by a linear combination of underlying risk drivers plus an additional return, i.e.
\begin{align}\label{eqn:asset}
r_t&=\alpha_t+\beta_tf_t+\varepsilon^r_t,&\varepsilon^r_t\sim\mathcal{N}(0,\Sigma_t^r),
\end{align}
for the present and all future times $t=T,\ldots,T+H$, up to forecasting horizon $H$. Here, $r_t$ is a vector of asset returns and $f_t$ is a vector of factor returns. The excess returns $\alpha_t$ and factor exposures $\beta_t$ are time-varying.\footnote{Factor exposures are also called factor loadings, or simply `beta'.} Equity factor exposures are typically estimated using regression, whereas for fixed-income assets the exposures are derived analytically at each future time through a rates model.\footnote{In practical settings it is more common to directly specify equity exposures, and derive the equity factor returns through linear regression \citep{sheikh1996barra}. The CSG is agnostic to this modeling choice as it assumes that exposures and factor returns are exogenous.} The error $\varepsilon^r_t$ represents idiosyncratic risk and is assumed to be independent across time and independent of all other sources of risk. Different assets may have correlated idiosyncratic risk in the sense that $\Sigma_t^r$ has non-zero off-diagonal elements.

The functional form of factor returns is slightly more general than is common in the literature, with means that can be time-varying,
\begin{align}\label{eqn:factor}
f_t&=\mu_t^f+\varepsilon^f_t,&\varepsilon^f_t\sim\mathcal{N}(0,\Sigma^f),
\end{align}
for all $t=1,\ldots,T+H$. Classical factor models are the Fama-French three-factor model \citep{fama1992,fama1993} for equity and the Nelson-Siegel model of the yield curve \citep{nelson1987} for fixed-income products.

The errors $\varepsilon^f_t$ are independent across time, but may be correlated with the errors $\varepsilon_t^x$ and $\varepsilon_t^y$ in the macroeconomic model.\footnote{Note that only the idiosyncratic risk $\varepsilon_t^r$ is uncorrelated with the other sources of risk identified thus far.} Without loss of generality, we may assume that $\varepsilon_t^f=F'\varepsilon_t'+F''\varepsilon_t''$, where $\varepsilon_t'$ and $\varepsilon_t''\sim\mathcal{N}(0,I)$ are independent sources of risk, and $F=\begin{bmatrix}F'&F''\end{bmatrix}$ is such that $FF^\top=\Sigma^f$.\footnote{Under certain assumptions, this structure of $F$ allows for separate estimation of the macroeconomic model and the factor model. Section \ref{sec:estimation} discusses this in more detail.}

In practical applications such as Markowitz portfolio optimization the interest is often in the distribution of returns conditional on the mean and covariance matrix. It follows from (\ref{eqn:asset}) and (\ref{eqn:factor}) that
\begin{alignat}{2}\label{eqn:asset2}
\mu_t&=&\E[r_t|\alpha_t,\mu_t^f]&=\alpha_t+\beta_t\mu_t^f,\\
\Sigma_t&=&\;\vcov[r_t|\alpha_t,\mu_t^f]&=\beta_t\Sigma^f\beta_t^\top+\Sigma_t^r.
\end{alignat}

The vector $\alpha_t$ that describes the additional return in excess of the factor model is common in the literature, but less is known about its behavior. I allow the possibility of non-zero alpha by assuming a mean-reverting stochastic process of the form
\begin{align}\label{eqn:alpha}
\alpha_{t+1}&=\Phi\alpha_t+\varepsilon^\alpha_t,
&\alpha_T&\sim\mathcal{N}(0,\tau\Sigma^r_T),
&\varepsilon^\alpha_t&\sim\mathcal{N}\left(0,\tau\Sigma_t^r-\tau\Phi\Sigma_t^r\Phi^\top\right),
\end{align}
for $t=T,\ldots,T+H$, with $\Phi$ diagonal or simply a constant. The errors $\varepsilon^\alpha_t$ are distributed independently across time and independent of all other variables in the framework. The covariance of $\varepsilon_t^\alpha$ is such that in case of homogeneity, i.e.~$\Sigma^r=\Sigma_t^r$, we obtain the unconditional distribution $\alpha_t\sim\mathcal{N}(0,\tau\Sigma^r)$. Therefore $\tau$ controls the tightness of the $\alpha_t$ process around zero (to be discussed in more detail below). For convenience, also introduce $\varepsilon_t'''\sim\mathcal{N}(0,I)$ such that $S_t\varepsilon_t'''\sim\varepsilon_t^\alpha$, where $S_t$ solves  $S_tS_t^\top=\tau\Sigma_t^r-\tau\Phi\Sigma_t^r\Phi^\top$. 

The AR(1) dynamics of each marginal alpha are consistent with \citet{mamaysky2008estimating}, who define alpha as the result of mean reverting trading signals. If an asset with constant positive alpha were to exist in excess of a sensible factor model, then given enough history investors would find it and invest in it. This would then increase the value of the asset and thereby diffuse its alpha. \citet{busse2010performance} find empirical evidence for the existence of alpha at shorter horizons for institutional investors. The speed of mean reversion and potential impact of alpha are encoded in $\Phi$ and $\tau$.\footnote{To ensure mean-reversion $\Phi$ should have values on the interval $(-1,1)$. The structure of the model allows stronger assumptions to be expressed, such as that a constant alpha vector $\alpha=\alpha_t$ exists. This alpha is unknown with prior $\alpha\sim\mathcal{N}(0,\tau\Sigma^r_T)$, and can be specified by the limit $\Phi\rightarrow I$ (such that $\varepsilon_t^\alpha=0$) and $\tau>0$. The constant alpha assumption has historically been the center of a large research agenda \citep[see for example][]{ferson1996measuring,barras2010false,fama2010luck}. Even stronger, the efficient market hypothesis states that $\alpha_t=0$ for any sensible choice of factor model. Choosing $\tau=0$ ($\Phi$ can be anything since $\alpha_T=0$ and $\varepsilon_t^\alpha=0$ as consequence) generates the dogmatic prior that there is no excess alpha.} The definition of $\tau$ in terms of the unconditional covariance matrix $\tau\Sigma^r$ may seem odd. I choose this structure because it uncovers a deep link with the parameter $\tau$ in the BL model and allows for a similar interpretation, as shown in Appendix \ref{sec:blacklitterman}. Intuitively, $\tau$ represents the tightness of the prior distribution of alpha around zero in the same way that $\tau$ defines the tightness of mean returns around the equilibrium in the BL model.

The tuple $\psi$ collects the parameters that describe the future markets (which can be defined independently of the views),
\begin{align}
\psi&=\langle \beta_t,\Sigma_t,\tau,\Phi\rangle_{t=T}^{T+H}.
\end{align}

\subsection{The macro-financial linkage}\label{sub:link}

As pointed out above, correlation between $\varepsilon_t^x$, $\varepsilon_t^y$ and $\varepsilon_t^f$ may exist, through which financial shocks can impact the economy and vice versa. This link is contemporaneous and therefore fast-moving and may not be useful for tactical asset allocations. 

The second way that a link between asset returns and macroeconomic variables can exist is through the mean of factor returns, $\mu_t^f$, using
\begin{align}\label{eqn:link}
\mu_t^f&=\bar{f}+\Gamma\tilde{x}_t.
\end{align}
This relationship reads that the \emph{mean} factor returns in excess of its steady state is linearly related to the latent macroeconomic states in excess of their respective steady states. Since the macroeconomic variables in $x_t$ are typically slow-moving, $\mu_t^f$ is also slow-moving. The matrix of loadings $\Gamma$ can describe typical stylized facts, e.g.~if GDP growth is higher than usual, then the return on the market factor also tends to be higher than usual.\footnote{This implies a potentially time-varying market price of risk for each factor. A non-zero loadings matrix $\Gamma$ is equivalent to saying that market risk premia are changing with the business cycle. As we are free to add lagged (or leading) versions of variables to the macroeconomic model, there may be an offset in the timing, in the sense that macroeconomic variables forecast risk premiums or the other way around. The fact that leading variables are not available for the latest time periods is not a problem, as the DLM framework handles missing values. If $y_t$ contains missing data at some time $t$, the corresponding rows in $y_t$, $B$ and $H$ can be removed. The resulting DLM is no longer time-homogeneous as $B$ and $H$ now vary through time, but all algorithms used in this paper accommodate this by default.}

There is ample theoretical and empirical literature on the existence of the link in (\ref{eqn:link}). \citet{cochrane2011presidential} outlines the basis of the theoretical argument. In a standard consumption-based model with power utility and log-normal consumption growth, the equity risk premium is a linear function of consumption growth and risk aversion.\footnote{Many richer structures can be identified by generalizing the framework. \citet{cochrane2011presidential} lists distinguishing durable and non-durable, traded or non-traded goods, as well as habit persistence, long-run risks and rare disasters. See \citet{claessens2018} and the references therein for a recent overview. For a more complete account, see \citet{campbell2003consumption}.}

Empirically, the macro-financial link has been studied for a wide array of factors. For the equity risk premium, the earliest proof came from dividend-price ratios and dividend yields. For example, \citet{campbell1988a,campbell1988b} show in two well-known papers that aggregate dividend yields forecast the mean of stock returns. Other variables that have been shown to have forecasting power are interest rate, spread and inflation related variables \citep{campbell1987stock,fama1989business,campbell2004inflation,ang2006}.

We do not strictly require a forecasting relationship between macroeconomic and financial variables. A contemporaneous effect, or even lagged relationship is sufficient for a non-zero $\Gamma$. Therefore, the relationship in (\ref{eqn:link}) is far more robust to the critiques outlined by \citet{welch2007comprehensive} that many existing equity return forecasting measures do not beat historical means out-of-sample. Also, for other asset classes the explanatory power of macroeconomic variables is far less controversial. For example, \citet{ang2003} find that up to 85\% of bond yields (i.e.~key rates) are explained by macroeconomic variables. \citet{chen1986economic} give an overviews of the kind of macroeconomic variables that may be considered for the right-hand side of (\ref{eqn:link}).

\subsection{Conditional forecasts}\label{sub:conditional}

The CSG framework describes the dynamics of $x_t$, $y_t$ and $f_t$ from time $t=1$ up to $T$. However, for the purpose of forecasting the interest is in the joint distribution of $x_t$, $y_t$, $f_t$ and $r_t$ from $t=T$ up to a forecasting horizon $t=T+H$. Moreover, this distribution should be conditional on (i.e.~consistently calibrated to) the future values of some of these variables, expressed as views $v_t$. To make this possible I assume that the macroeconomic model and macro-financial link remain valid up to time $T+H$, even if the observations end at $T$. In general all views can be combined in a single matrix equation of the following form,
\begin{align}
\tilde{v}_t&=P_t\alpha_t+Q_t\tilde{x}_t+R_t\varepsilon_t+\xi_t,&\xi_t&\sim\mathcal{N}(0,\Omega_t),
\end{align}
for $t=T,\ldots,T+H$, where $\varepsilon_t=(\varepsilon_t',\varepsilon_t'',\varepsilon_t''')$ are the macroeconomic, factor, and alpha-related independent sources of risk. $\xi_t$ is an additional source of risk that describes the uncertainty of the views. Views may be exact in the sense that they have no uncertainty, by choosing $\Omega_t$ as zero. This paper distinguishes several different types of conditioning that can be written in this format.
\begin{description}
\item[Views on macroeconomic states] are values to condition the future value of macroeconomic latent state variables on. These views are expressed through the matrix equation
\begin{align}
v^x_t&=Q_t^xx_t+\xi_t^x,&\xi_t^x&\sim\mathcal{N}(0,\Omega^x_t),\\
\tilde{v}_t^x&=v_t^x-Q_t^x\bar{x}=Q_t^x\tilde{x}_t+\xi_t^x\nonumber
\end{align}
where $v_t^x$ stores the views and $Q_t^x$ maps the views to the variables. For example, suppose the aim is to condition on GDP growth being $-2\%$ at $t=T+5$ with $1\%$ standard deviation, and GDP growth is stored in the second entry in $x_t$, then $v_{T+5}=-0.02$, $Q_{T+5}=\begin{bmatrix}0&1&0&\cdots&0\end{bmatrix}$ and $\Omega^x_{T+5}=0.01^2$. Adding additional views to the same time $t$ expands the rows of $v_t^x$ and $Q_t^x$ and the rows and columns of $\Omega^x_t$.
\item[Views on macroeconomic observations] are values to condition the future value of macroeconomic observable time-series on,
\begin{align}
v^y_t&=Q_t^yy_t+\xi_t^y=Q_t^y(\bar{y}+B\tilde{x}_t+\varepsilon_t^y)+\xi_t^y,&\xi_t^y&\sim\mathcal{N}(0,\Omega^y_t),\\
\tilde{v}^y_t&=v_t^y-Q_t^y\bar{y}=Q_t^yB\tilde{x}_t+Q_t^yH\varepsilon_t'+\xi_t^y.\nonumber
\end{align}
The interpretation of the components is the same as above. Views on $y_t$ are important because the latent processes $x_t$ are not always meaningful to condition on. For example, in the case of a FAVAR macroeconomic model the values in $x_t$ are estimated via principal component analysis, whereas the observable variables $y_t$ are interpretable. 
\item[Views on macroeconomic shocks] are values to condition the shocks to the macroeconomic system on,
\begin{align}
v^\varepsilon_t&=R_t^\varepsilon\varepsilon_t'+\xi_t^\varepsilon,&\xi_t^\varepsilon&\sim\mathcal{N}(0,\Omega_t^\varepsilon),\\
\tilde{v}^\varepsilon_t&=v_t^\varepsilon=R_t^\varepsilon\varepsilon_t'+\xi_t^\varepsilon.\nonumber
\end{align}
In macroeconomic theory these views are important because they can be used to create impulse response functions. These are the responses of a system that is in steady state to a single-period view on exactly one element of $\varepsilon_t'$. Since the interpretation of the elements of $\varepsilon_t'$ depends on $G$ and $H$, an identification problem arises for this specific type of views. Section \ref{sub:macro} gives some references for dealing with this issue.
\item[Views on mean factor returns] are values to condition the future mean factor returns on,
\begin{align}
v^{\mu_f}_t&=P_t^{\mu_f}\mu^f_t+\xi_t^{\mu_f}=P_t^{\mu_f}(\bar{f}+\Gamma\tilde{x}_t)+\xi_t^{\mu_f},&\xi_t^{\mu_f}&\sim\mathcal{N}(0,\Omega_t^{\mu_f}),\\
\tilde{v}^{\mu_f}_t&=v_t^{\mu_f}-P_t^{\mu_f}\bar{f}=P_t^{\mu_f}\Gamma\tilde{x}_t+\xi_t^{\mu_f}.\nonumber
\end{align}
The interpretation is similar. Views on mean factor returns make sense from an investment perspective, for example when modeling a financial crisis or when simply adjusting the model forecasts with investor views such as CMAs.
\item[Views on factor returns] are values to condition the future factor returns on,
\begin{align}
v^f_t&=P_t^ff_t+\xi_t^f=P_t^f(\bar{f}+\Gamma\tilde{x}_t+\varepsilon_t^f)+\xi_t^f,&\xi_t^f&\sim\mathcal{N}(0,\Omega_t^f),\\
\tilde{v}^f_t&=v_t^f-P_t^f\bar{f}=P_t^f\Gamma\tilde{x}_t+P_t^fF'\varepsilon_t'+P_t^fF''\varepsilon_t''+\xi_t^f.\nonumber
\end{align}
When we want to condition on actual factor returns instead of mean factor returns, we can modify the structure slightly to add the error in the factor equation (\ref{eqn:factor}).
\item[Views on mean asset returns] are values to condition the future mean asset returns on,
\begin{align}
v^\mu_t&=P_t^\mu\mu_t+\xi_t^\mu=P_t^f(\alpha_t+\beta_t(\bar{f}+\Gamma\tilde{x}_t))+\xi_t^\mu,&\xi_t^\mu&\sim\mathcal{N}(0,\Omega_t^\mu),\\
\tilde{v}^\mu_t&=v_t^\mu-P_t^\mu\beta_t\bar{f}=P_t^\mu\alpha_t+P_t^\mu\beta_t\Gamma\tilde{x}_t+\xi_t^\mu.\nonumber
\end{align}
Views on asset mean returns are useful when the interest is in returns on specific assets, and when at the same time analyst forecasts are available for these specific stocks. These forecasts can be assimilated by the model in the form of views.
\end{description}

Conditional on the views $v_t$ at future times $t=T,\ldots,T+H$ it is possible to generate the future means, covariances and paths of the macroeconomic variables $x_t$, the measurements $y_t$, and all factor and asset returns and mean returns. Mathematically, what we want to forecast or simulate is
\begin{align}
x_t,\alpha_t,\varepsilon_t&|y_1,\ldots,y_T,f_1,\ldots,f_T,v_T,\ldots v_{T+H},&t&=T,\ldots,T+H.
\end{align}
All other variables of interest are linear combinations of these variables. Appendix \ref{sec:forecasting} describes the approach to generate these forecasts in more detail. This approach allows us to forecast and simulate linearly in the number of time-steps, and cubically in the number of variables. Note that this is the same computational complexity as a standard Monte-Carlo simulation without conditioning, if there is a time-inhomogeneous correlation structure in the variables that requires factorization.\footnote{To simulate from jointly normal random variables at every time step, a Cholesky or LDL decomposition is required that runs in cubic time.}

The tuple $\phi$ collects the parameters required to describe the mapping of views and their uncertainty,
\begin{align}
\phi&=\langle P_t,Q_t,R_t,\Omega_t\rangle_{t=T}^{T+H}.
\end{align}
Not all future times may have views, in which case the corresponding matrices have zero rows, and also zero columns for $\Omega_t$.

\section{Estimation procedures}\label{sec:estimation}

By aggregating (\ref{eqn:dlm1}-\ref{eqn:dlm2}), (\ref{eqn:factor}), and (\ref{eqn:link}) into a single DLM, we can obtain
\begin{align}
\label{eqn:dlm3}
\tilde{x}_{t+1}&=A\tilde{x}_t+\begin{bmatrix}G&O\end{bmatrix}\begin{bmatrix}\varepsilon_t'\\\varepsilon_t''\end{bmatrix},
&\tilde{x}_1&\sim\mathcal{N}(\tilde{x}_{1|0},P_{1|0}),\\
\label{eqn:dlm4}
\begin{bmatrix}\tilde{y}_t\\\tilde{f}_t\end{bmatrix}
&=\begin{bmatrix}B\\\Gamma\end{bmatrix}\tilde{x}_t+\begin{bmatrix}H&O\\F'&F''\end{bmatrix}\begin{bmatrix}\varepsilon_t'\\\varepsilon_t''\end{bmatrix},&\begin{bmatrix}\varepsilon_t'\\\varepsilon_t''\end{bmatrix}&\sim\mathcal{N}(0,I),
\end{align}
where $F=\begin{bmatrix}F'&F''\end{bmatrix}$ is such that $FF^\top=\Sigma^f$ and can be obtained from a Cholesky decomposition of the joint covariance matrix of $(\varepsilon'_t, \varepsilon_t^f)$, and $O$ denotes a zero matrix of appropriate size. The asset returns described in (\ref{eqn:asset}) and alphas in (\ref{eqn:alpha}) are not added because the errors therein, $\varepsilon_t^r$ and $\varepsilon^\alpha_t$, are independent of $\varepsilon^x_t$, $\varepsilon^y_t$ and $\varepsilon_t^f$. Note how the only structural change to the DLM in (\ref{eqn:dlm1}-\ref{eqn:dlm2}) is the additional measurements of the macroeconomic states, and additional sources of risk in the measurement errors. 

Conceptually, there are two ways to estimate this model, regardless of whether we pick a frequentist or Bayesian perspective. The first approach is a full re-estimation of the macroeconomic model with the new measurements $f_t$, based on the idea that the high-level structure of the macroeconomic model remains the same. Only new measurement equations have been added, but these may influence the matrices in the macroeconomic model. For some models such as the VAR class re-estimation may be straightforward, for DSGE models this is harder. The second approach is to re-use the original estimation of the macroeconomic model, and estimate the macro-financial link separately. This requires an additional assumption, namely that factor returns contain no information on the parameters and variables of the macroeconomic model, given the historical observations. The next sections explain these two approaches in more detail.

\subsection{Estimating macro and financial models jointly}\label{sub:estjointly}

As pointed out, full re-estimation based on additional observations is simple for VAR-type models. For example, in a standard VAR all latent states are observed so $x_t=y_t$ and thus $H=O$ and $B=I$. The fact that $x_t$ are observable in this setting allows us to estimate $B$ and $\Gamma$ using seemingly unrelated regressions (SUR). There are only a few more equations to run. The matrices $G$ and $F$ can subsequently be obtained by Cholesky decomposition on the sample covariance of the residuals.

For VAR models with latent states, such as the FAVAR approach, we can simply assume that the factors are additional observations. Any FAVAR is constructed from a large number of time-series, so the methodology allows for additional series without modification \citep{bernanke2005measuring}.

DSGE models are trickier. These models are typically estimated using MCMC methods. The size of $\Gamma$ and $F$ can cause the number of parameters to grow rapidly with the number of factors, rendering MCMC less feasible. \citet{boivin2006dsge} propose a solution, by inserting a Gibbs sampling step inside the MCMC. The algorithm below is a straightforward modification using the present notation.\footnote{In the specific case that we already start out with a formulation as in \citet{boivin2006dsge}, i.e.~$B$ and $H$ are not functions of a small set of underlying parameters, but need to be estimated in full, no modifications are required and we can view the factors as additional measurements as in the FAVAR approach.}

Using a solver such as the algorithm by \citet{anderson1985linear}, given a set of parameters $\pi$ that calibrates the macroeconomic model, we can write the DLM in (\ref{eqn:dlm3}-\ref{eqn:dlm4}) as follows.
\begin{align}
\label{eqn:dlm5}
\tilde{x}_{t+1}&=A(\pi)\tilde{x}_t+\begin{bmatrix}G(\pi)&O\end{bmatrix}\begin{bmatrix}\varepsilon_t'\\\varepsilon_t''\end{bmatrix},&\tilde{x}_1&\sim\mathcal{N}(\tilde{x}_{1|0},P_{1|0}),\\
\label{eqn:dlm6}
\begin{bmatrix}\tilde{y}_t\\\tilde{f}_t\end{bmatrix}&=\begin{bmatrix}B(\pi)\\\Gamma\end{bmatrix}\tilde{x}_t+\begin{bmatrix}H(\pi)&O\\F'&F''\end{bmatrix}\begin{bmatrix}\varepsilon_t'\\\varepsilon_t''\end{bmatrix},&\begin{bmatrix}\varepsilon_t'\\\varepsilon_t''\end{bmatrix}&\sim\mathcal{N}(0,I),
\end{align}
where the steady states $\bar{y}$ and $\bar{x}$ can also be functions of $\pi$, but the vector of factor means $\bar{f}$ is not. With some initial parameter draw $\pi^{(0)}$, $\Gamma^{(0)}$, $F^{(0)}$ and $\bar{f}^{(0)}$, iterate through the following steps.
\begin{enumerate}
    \item Draw the latent time-series given the parameters and data, 
    \begin{align*}
    p\big(x_{1:T}^{(i)},
    \big|\pi^{(i-1)},\Gamma^{(i-1)},F^{(i-1)},\bar{f}^{(i-1)},y_{1:T},f_{1:T}\big). 
    \end{align*}
    This is done using a standard simulation smoother, such as described by \citet[Appendix A]{durbin2002simple}.
    \item Draw the linear parameters given the parameters $\pi$, the latent time-series and the data,
    \begin{align*}
    p\big(\Gamma^{(i)},F^{(i)},\bar{f}^{(i)}\big|\pi^{(i-1)},x_{1:T}^{(i)},
    y_{1:T},f_{1:T}\big). 
    \end{align*}
    For example, with a normal-inverse-Wishart conjugate prior standard procedures can be used to sample this distribution.
    \item Draw the parameters $\pi$ given the linear parameters, the latent time-series and the data,
    \begin{align*}
    p\big(\pi^{(i)} \big|\Gamma^{(i)},F^{(i)},\bar{f}^{(i)},x_{1:T}^{(i)},
    y_{1:T},f_{1:T}\big). 
    \end{align*}
    Due to the non-linearity, we require a likelihood-based accept-reject step here.
\end{enumerate}

\subsection{Estimation of macroeconomic model first}\label{sub:estmacrofirst}

I now consider estimation of the framework when factor returns contain no information for the estimation of the macroeconomic model. In a Bayesian setting, this can be expressed as the following conditional independence,
\begin{align*}
    p(\theta_m,x_{1:T},\varepsilon'_{1:T}|y_{1:T},f_{1:T})=p(\theta_m,x_{1:T},\varepsilon'_{1:T}|y_{1:T}).
\end{align*}
where the tuple $\theta$ collects all parameters to be estimated, i.e.
\begin{align}
\theta&=\langle A,B,\Gamma,G,H,F,\bar{x},\bar{y},\bar{f}\rangle
=\langle\theta_m,\theta_f\rangle,
&\theta_m&=\langle A,B,G,H,\bar{x},\bar{y}\rangle,
&\theta_f&=\langle\Gamma,F,\bar{f}\rangle,
\end{align}
with $\theta_m$ the parameters specific to the macroeconomic model, and $\theta_f$ the macro-financial link. With this assumption in place, we can split the estimation using Bayes rule,
\begin{align*}
    p(\theta,x_{1:T},\varepsilon'_{1:T},\varepsilon''_{1:T}|y_{1:T},f_{1:T})
    &=p(\theta_f,\varepsilon''_{1:T}|x_{1:T},\varepsilon'_{1:T},y_{1:T},f_{1:T})p(\theta_m,x_{1:T},\varepsilon'_{1:T}|y_{1:T},f_{1:T})\\
    &=p(\theta_f,\varepsilon''_{1:T}|x_{1:T},\varepsilon'_{1:T},y_{1:T},f_{1:T})p(\theta_m,x_{1:T},\varepsilon'_{1:T}|y_{1:T}).
\end{align*}

For a Bayesian estimation of the DLM in (\ref{eqn:dlm5}-\ref{eqn:dlm6}) that uses this assumption, I propose the following two-step estimation procedure. 
\begin{enumerate}
\item Using whatever method is available to the macroeconomic model, we draw from the posterior distribution of $p(\theta_m,x_{1:T},\varepsilon'_{1:T}|y_{1:T})$. In case we only have a sample from the posterior parameters $p(\pi|y_{1:T})$ available, then we can use the mapping from $\pi$ to $\theta_m$ that is implicit in (\ref{eqn:dlm5}-\ref{eqn:dlm6}), as well as a standard simulation smoother to generate this sample.
\item We draw from the distribution $p(\theta_f|\tilde{x}_{1:T},f_{1:T},\varepsilon'_{1:T})$. This second step is a Bayesian multivariate linear regression with explanatory variables $\tilde{x}_t$ and $\varepsilon'_t$ as well as an intercept. I.e.
\begin{align*}
f_t&=\bar{f}+\Gamma\tilde{x}_t+F'\varepsilon'_t+F''\varepsilon''_t.
\end{align*}
For notational convenience, I write the regression in this second step as
\begin{align*}
\mathcal{Y}=\mathcal{X}\mathcal{B}+\mathcal{E},
\end{align*}
with $\mathcal{Y}=\tilde{f}_{1:T}^\top$, $\mathcal{X}=\begin{bmatrix}1&\tilde{x}_{1:T}^\top&\varepsilon_{1:T}'^\top\end{bmatrix}$, $\mathcal{B}=\begin{bmatrix}\bar{f}&\Gamma&F'\end{bmatrix}$ and $\mathcal{E}=(F''\varepsilon''_{1:T})^\top$. Also denote $\Sigma_{\mathcal{E}}=F''F''^\top$, which is the covariance matrix of the rows of $\mathcal{E}$.
\end{enumerate}
Notice how the estimation of the macro and factor part are separated. We do not need to adjust the estimation procedure of the macroeconomic model to the added factor block.

For the sake of completeness and because Section \ref{sec:egC} implements this specific setup, I will given an example with a flat normal-inverse-Wishart conjugate prior for the parameters $\mathcal{B}$ and $\Sigma_{\mathcal{E}}$. This means that the covariance matrix of $\Sigma_{\mathcal{E}}$ is inverse-Wishart distributed, and conditional on this covariance matrix the coefficients $\mathcal{B}$ follow the matrix-normal distribution. That is,
\begin{align*}
(\Sigma_{\mathcal{E}}|\mathcal{Y},\mathcal{X})
&\sim\mathcal{W}^{-1}(\hat{V}_0,\hat{\nu}_0)
\\
(\mathcal{B}|\mathcal{Y},\mathcal{X},\Sigma_{\mathcal{E}})
&\sim\mathcal{MN}(\hat{B}_0,\hat{\Lambda}_0^{-1},\Sigma_{\mathcal{E}}),
\end{align*}
where $\hat{V}_0$, $\hat{\nu}_0$, $\hat{B}_0$ and $\hat{\Lambda}_0$ are parameters controlling the prior.

The updating formulas follow from the standard formulas for Bayesian multivariate linear regression \citep{karlsson2013forecasting}. The posterior parameters are
\begin{align*}
\hat{\Lambda}&=\hat{\Lambda}_0+\mathcal{X}^\top\mathcal{X},\\
\hat{B}&=\hat{\Lambda}^{-1}(\mathcal{X}^\top\mathcal{Y}+\hat{\Lambda}_0\hat{B}_0),\\
\hat{\nu}&=\hat{\nu}_0+T,\\
\hat{V}&=\hat{V}_0+(\mathcal{Y}-\mathcal{X}\hat{B})^\top(\mathcal{Y}-\mathcal{X}\hat{B})+(\hat{B}-\hat{B}_0)^\top\hat{\Lambda}_0(\hat{B}-\hat{B}_0).
\end{align*}
For a flat prior, we have $\hat{V}_0=O$, $\hat{\nu}_0=n_f-n_x-n_{\varepsilon'}$, $\hat{B}_0=O$ and $\hat{\Lambda}_0=O$, such that the estimation reduces to OLS, which we can sample from using the normal-inverse-Wishart distribution.

\section{Example A: FAVAR with Fama-French and Nelson-Siegel factors}\label{sec:egB}

This and the next section give two example applications of the CSG. For the first example I choose an empirical macroeconomic model. The FAVAR model of \citet{bernanke2005measuring} identifies a number of latent factors that drive a larger number of macroeconomic time-series. The model can be estimated using principal-component analysis (PCA).\footnote{The authors also implement a Gibbs sampler, but as these methods give very similar results, the simpler PCA-based approach is taken here.} The factors come from the \citet{fama1992, fama1993} three factor (FF3) model constructed from US stock returns data, and I use the Nelson-Siegel \citeyearpar{nelson1987} model to explain the US treasury yield curve with a level, a slope and a curvature factor. This brings the total to six factors.

\subsection{Methodology}\label{sec:egBmethodology}

This subsection discusses the methodology behind the macroeconomic model, the financial markets model, and the macro-financial link. I use the estimation approach explained in Section \ref{sub:estmacrofirst}, i.e.~to estimate the macroeconomic model first, under the assumption that the factor returns provide no additional information. The conditional forecasts are formulated as outlined in Section \ref{sub:conditional}. I use Appendix \ref{sec:forecasting} to produce the analytical conditional distributions.

\subsubsection{Macroeconomic model}

The details of estimating the FAVAR model are quite involved, and I refer to the original paper for the exact PCA-based method. The estimated model can be written in the form of (\ref{eqn:dlm1}-\ref{eqn:dlm2}). Here $x_t$ holds the federal funds rate, five latent drivers of the economy and six lags of each of these six variables.\footnote{In the paper, the latent variables form a VAR(7), but we only need additional states for lags beyond the first, hence the inclusion of six lags.} The measurement variables $y_t$ are 120 macroeconomic time-series, including the federal funds rate (details in Section \ref{sec:egBdata}). The estimation procedure gives an estimate of the tuple $\theta_m$, namely $A$, $B$, $G$, $H$, $\bar{x}$ and $\bar{y}$. Additionally, the PCA-based approach returns estimates of the latent drivers stored in $x_t$.

\subsubsection{Financial markets model}

I use the canonical three Fama-French factors, and include the Nelson-Siegel factors to describe the yield curve. A brief explanation of this model follows.

In the Nelson-Siegel framework, the yield curve is explained by three factors: level $f^{\mathrm{L}}_t$, slope $f^{\mathrm{S}}_t$ and curvature $f^{\mathrm{C}}_t$, jointly denoted $f^{\mathrm{LSC}}_t=(f^{\mathrm{L}}_t,f^{\mathrm{S}}_t,f^{\mathrm{C}}_t)$. Let $P(t,T)$ be the price of a zero-coupon bond with maturity $T$ at time $t$. Then the $T$-yield at time $t$, $R(t,T)$, is defined through
\begin{align*}
    P(t,T)&=\exp{\{-R(t,T)(T-t)\}},\\
    R(t,T)&=\frac{1}{T-t}\ln P(t,T)=B_{\mathrm{L}}(T-t,\lambda)f^{\mathrm{L}}_t+B_{\mathrm{S}}(T-t,\lambda)f^{\mathrm{S}}_t+B_{\mathrm{C}}(T-t,\lambda)f^{\mathrm{C}}_t,
\end{align*}
where
\begin{align*}
B_{\mathrm{L}}(\tau,\lambda)&=1,&B_{\mathrm{S}}(\tau,\lambda)&=\frac{1-\exp{\{-\lambda\tau\}}}{\lambda\tau},&B_{\mathrm{C}}(\tau,\lambda)&=\frac{1-\exp{\{-\lambda\tau\}}}{\lambda\tau}-\exp{\{-\lambda\tau\}}.
\end{align*}
Clearly, the log-price of the bond is linear in the factors. Since the marginal distributions through time of the factors are normal, the marginal distributions of the bond price are log-normal with an analytical confidence interval.\footnote{It is easy to see that the log-return on the bond is linear in $(f^{\mathrm{L}}_t,f^{\mathrm{S}}_t,f^{\mathrm{C}}_t,f^{\mathrm{L}}_{t-1},f^{\mathrm{S}}_{t-1},f^{\mathrm{C}}_{t-1})$. Although the lags of the factors are not included in the model, they can be added by extending the DLM with the lags of the corresponding errors. I choose to model the price here to show how the pull-to-par effect is preserved in the framework.}

For known $\lambda$, the factors can be estimated using linear regression at each time $t$. For a set of rates with maturities $\tau_j$, $j=1,\ldots,k$ that is available at each time $t$, we can run the cross-sectional regressions
\begin{align*}
\begin{bmatrix}R(t,t+\tau_1)\\\vdots\\R(t,t+\tau_k)\end{bmatrix}
=\underbrace{\begin{bmatrix}B_{\mathrm{L}}(\tau_1,\lambda)&B_{\mathrm{S}}(\tau_1,\lambda)&B_{\mathrm{C}}(\tau_1,\lambda)\\\vdots&\vdots&\vdots\\B_{\mathrm{L}}(\tau_k,\lambda)&B_{\mathrm{S}}(\tau_k,\lambda)&B_{\mathrm{C}}(\tau_k,\lambda)\end{bmatrix}}_{X(\lambda)}
\begin{bmatrix}f^{\mathrm{L}}_t\\f^{\mathrm{S}}_t\\f^{\mathrm{C}}_t\end{bmatrix}+\begin{bmatrix}\eta_{1,t}\\\vdots\\\eta_{k,t}\end{bmatrix},
\end{align*}
with all errors $\eta_{j,t}$ i.i.d.~across tenors and time. I apply non-linear least-squares on all parameters $(\lambda,f^{\mathrm{L}}_1,f^{\mathrm{S}}_1,f^{\mathrm{C}}_1,\ldots,f^{\mathrm{L}}_T,f^{\mathrm{S}}_T,f^{\mathrm{C}}_T)$ by a grid search over $\lambda$ and running all cross-sectional least-squares minimizations for the independent variables $X(\lambda)$. 

\subsubsection{Macro-financial link}

There are five tenors included in the FAVAR data-series $y_t$ by default, namely, the 3-and 6-month and 1, 5-and 10-year treasury rates. After estimating $\hat{X}=X(\hat{\lambda})$, we can obtain the level, slope and curvature factors from $y_t$ by pre-multiplying the subset of $y_t$ that contains the rates, $y_{\mathcal{R},t}$, with $(\hat{X}^\top\hat{X})^{-1}\hat{X}^\top$. That is,
\begin{align*}
f^{\mathrm{LSC}}_t&=(\hat{X}^\top\hat{X})^{-1}\hat{X}^\top y_{\mathcal{R},t}
=(\hat{X}^\top\hat{X})^{-1}\hat{X}^\top(B_{\mathcal{R}}x_t+H_{\mathcal{R}}\varepsilon'_t),
\end{align*}
where the subscript $\mathcal{R}$ again indicates that we are dealing with the rows corresponding to the rate observations. When we add the Fama-French factors $f^{\mathrm{FF3}}_t=(f^{\mathrm{SMB}}_t,f^{\mathrm{HML}}_t,f^{\mathrm{LSC}}_t)$ and their sources of risk $\varepsilon_t''$, we get the macro-financial link in the format of (\ref{eqn:dlm4}),
\begin{align*}
f_t&=\begin{bmatrix}f_t^{\mathrm{FF3}}\\f_t^{\mathrm{LSC}}\end{bmatrix}
=\underbrace{\begin{bmatrix}\bar{f}^{\mathrm{FF3}}-\Gamma_{\mathcal{F}}\bar{x}\\0\end{bmatrix}}_{\bar{f}-\Gamma\bar{x}}
+\underbrace{\begin{bmatrix}\Gamma_{\mathcal{F}}\\(\hat{X}^\top\hat{X})^{-1}\hat{X}^\top B_{\mathcal{R}}\end{bmatrix}}_{\Gamma}x_t
+\underbrace{\begin{bmatrix}F_{\mathcal{F}}'&F_{\mathcal{F}}''\\(\hat{X}^\top\hat{X})^{-1}\hat{X}^\top H_{\mathcal{R}}&O\end{bmatrix}}_{F}
\begin{bmatrix}\varepsilon_t'\\\varepsilon_t''\end{bmatrix},
\end{align*}
where the subscript $\mathcal{F}$ takes the rows corresponding to the Fama-French factors. $\Gamma$ and $F$ can be estimated block-wise, by estimating $\Gamma_\mathcal{F}$ and $F_\mathcal{F}$ through regression of the Fama-French factors on the estimated states $x_t$, and using the estimates of $B$ and $H$ to construct the lower blocks. However, this turns out to be equivalent to estimating $\Gamma$ and $F$ directly by running regressions of $f_t$ on $x_t$.

\subsection{Data}\label{sec:egBdata}

To estimate the FAVAR model, I use the same data as \citet{bernanke2005measuring}, i.e.~120 macroeconomic series on (A) real output and income, (B) employment and hours, (C) consumption, (D) housing starts and sales, (E) real inventories, orders, and unfilled orders, (F) stock prices, (G) exchange rates, (H) interest rates, (I) money and credit quantity aggregates, (J) prices indexes, and (K) average hourly earnings. All series have history from January 1959 through August 2001.\footnote{I have chosen not to update the data series with more recent data, since several of the series have been retired since, and the focus of this section is illustration, not prediction.}

Since the Nelson-Siegel factors are estimated from the same data, the factor returns for the FF3 model for the US market need to be added \citep{french2019data}. All arithmetic returns are transformed to annualized log-returns. The data contains the risk free rate (RFR), the market return in excess of the risk free rate (MKT), the returns on a portfolio long in small stocks and short in big stocks measured by market capitalization (small minus big, SMB), and a similar portfolio long in high book to value stocks and short low book to value stocks (high minus low, HML). The FF3 factor model is formulated in excess of the risk free rate, but can be rewritten in terms of total returns by adding the RFR as a factor that all equity has unit exposure to.

\subsection{Results}\label{sec:egBresults}

This section gives the estimation results and compares unconditional and conditional forecast. For the conditional part, I use a scenario where the price of a 5-year zero-coupon bond with face value \$100, purchased for \$80 at the time the prediction starts (implied by the 4.5\% 5-year rate at August 2001), is worth \$85 at the 3-year horizon. This intuitively is a reverse scenario analysis: we want to see what kind of macroeconomic scenario we need to meet an unhedged liability in the future.

The number of parameters estimated in the FAVAR is too large to display here efficiently, but a partial analysis is available in \citet{bernanke2005measuring}. Table \ref{tbl:favarest} describes how the factors are explained by the latent drivers in the macroeconomic model. By and large there is a fairly strong link between the market factor and the FAVAR factors. The Nelson-Siegel factors show an even stronger link, as is to be expected from the inclusion of various rates in the FAVAR model.

\begin{table}[!ht] \centering 
  \caption{Parameter estimates for the macro-financial link in the FAVAR example.} 
  \label{tbl:favarest} 
\begin{tabular}{@{\extracolsep{-15pt}}lD{.}{.}{-3} D{.}{.}{-3} D{.}{.}{-3} D{.}{.}{-3} D{.}{.}{-3} D{.}{.}{-3} } 
\\[-1.8ex]\hline 
\hline \\[-1.8ex] 
\\[-1.8ex] & \multicolumn{6}{c}{\textit{Macro-financial linkage:}} \\ 
 & \multicolumn{1}{c}{MKT} & \multicolumn{1}{c}{SMB} & \multicolumn{1}{c}{HML} & \multicolumn{1}{c}{level} & \multicolumn{1}{c}{slope} & \multicolumn{1}{c}{curvature} \\ 
\hline \\[-1.8ex] 
 FYFF & -1.569^{*} & -1.039 & 0.345 & 5.535^{***} & 4.188^{***} & 3.452^{***} \\ 
  & (0.884) & (0.676) & (0.625) & (0.141) & (0.160) & (0.425) \\ 
  PC1 & -0.718 & 0.566 & 0.065 & 1.229^{***} & -1.510^{***} & -0.551 \\ 
  & (0.812) & (0.622) & (0.575) & (0.129) & (0.147) & (0.390) \\ 
  PC2 & -4.116^{***} & -0.101 & 0.383 & -0.523^{***} & 0.456^{**} & -0.214 \\ 
  & (0.990) & (0.758) & (0.700) & (0.157) & (0.180) & (0.476) \\ 
  PC3 & -1.165 & -0.913 & -0.159 & -4.721^{***} & 4.337^{***} & -1.324^{***} \\ 
  & (0.762) & (0.583) & (0.539) & (0.121) & (0.138) & (0.366) \\ 
  PC4 & -4.696^{***} & -1.491^{**} & 1.011 & -1.452^{***} & 1.534^{***} & 0.388 \\ 
  & (0.931) & (0.713) & (0.659) & (0.148) & (0.169) & (0.447) \\ 
  PC5 & 7.025^{***} & 4.528^{***} & -1.550^{**} & -1.842^{***} & 1.935^{***} & 0.380 \\ 
  & (0.984) & (0.753) & (0.696) & (0.156) & (0.178) & (0.473) \\ 
  Const. & 16.090^{***} & 8.839^{*} & 3.110 & 48.505^{***} & -42.640^{***} & 2.670 \\ 
  & (6.139) & (4.698) & (4.344) & (0.976) & (1.114) & (2.949) \\ 
 \hline \\[-1.8ex] 
$N$ & \multicolumn{1}{c}{504} & \multicolumn{1}{c}{504} & \multicolumn{1}{c}{504} & \multicolumn{1}{c}{504} & \multicolumn{1}{c}{504} & \multicolumn{1}{c}{504} \\ 
R$^{2}$ & \multicolumn{1}{c}{0.191} & \multicolumn{1}{c}{0.085} & \multicolumn{1}{c}{0.016} & \multicolumn{1}{c}{0.927} & \multicolumn{1}{c}{0.816} & \multicolumn{1}{c}{0.205} \\ 
$\sigma_\varepsilon$ & \multicolumn{1}{c}{48.281} & \multicolumn{1}{c}{36.952} & \multicolumn{1}{c}{34.164} & \multicolumn{1}{c}{7.677} & \multicolumn{1}{c}{8.759} & \multicolumn{1}{c}{23.192} \\ 
\hline 
\hline \\[-1.8ex] 
\textit{Note:}  & \multicolumn{6}{r}{$^{*}$p$<$0.1; $^{**}$p$<$0.05; $^{***}$p$<$0.01} \\ 
\end{tabular} 
\end{table} 

\begin{figure}[!ht]
\centering
\includegraphics[width=1\textwidth]{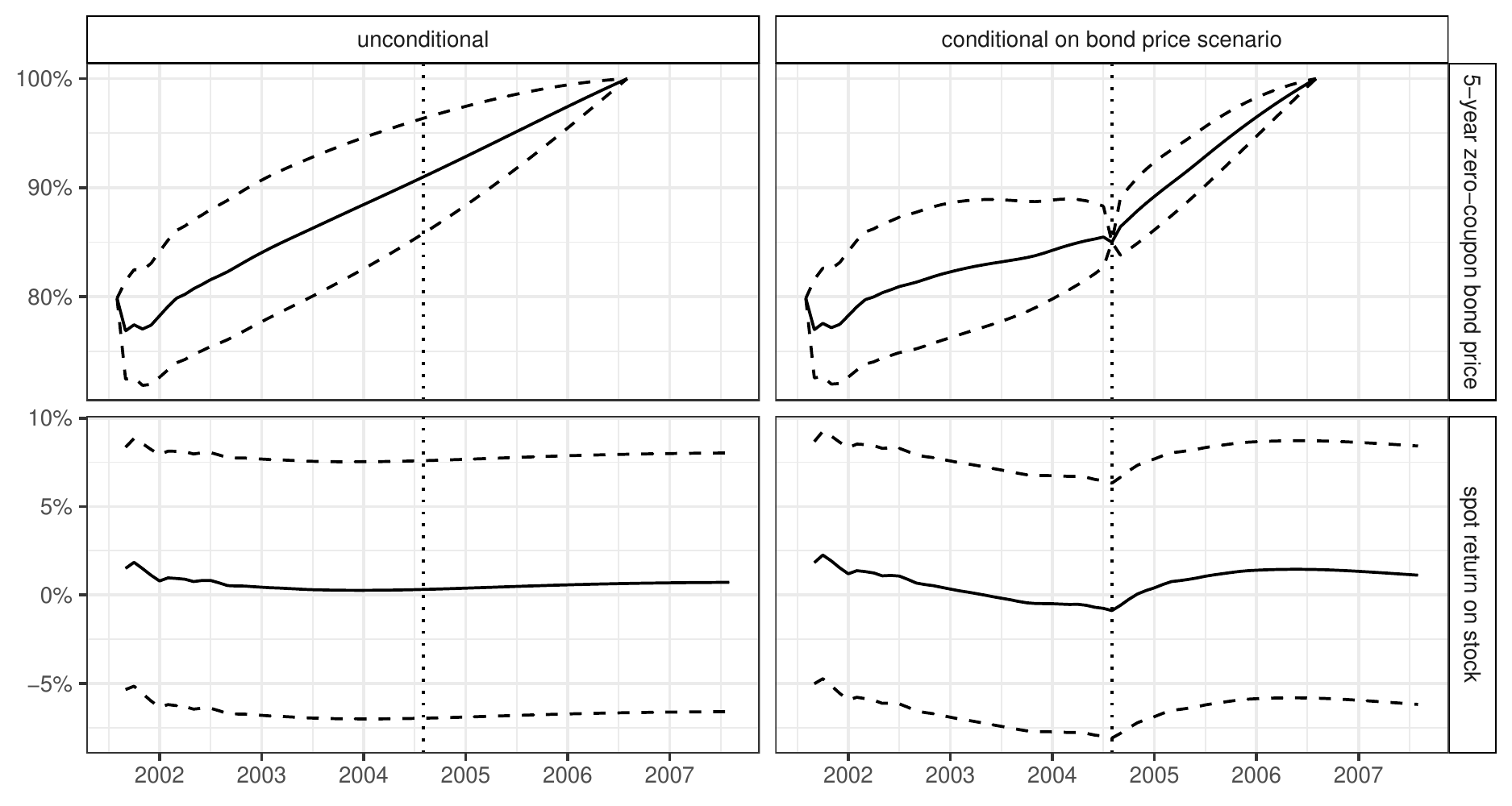}
\caption{Unconditional versus conditional (columns) predictions of the asset-level predictions (rows). The thick line plots the mean forecast. The dashed lines indicate a 90\% confidence interval. The timing of the conditional view is indicated by the vertical dotted line, and as the view is exact (no uncertainty), the confidence interval width of the bond price shrinks to zero as the bond matures and its value pulls to par.}
\label{fig:exampleBasset}
\end{figure}

Because the conditional problem is set up as a reverse stress test, I start with the graphs for the assets. Next to the price of the zero coupon bond, this includes a stock with unit exposure to the short-rate (using exposures to the Nelson-Siegel factors), and unit exposure to the market factor. This asset has an annualized excess return variance of 15\%. Figure \ref{fig:exampleBasset} shows the price evolution of the bond as well as the spot return on the stock. From the unconditional case, it is evident that the scenario is roughly the lower 5th percentile of the bond price projection. By restricting on this price, the confidence interval shrinks to zero at the 3-year horizon in the conditional case. We also see a strong response from the asset, with a dip around the same horizon and a subsequent recovery.\footnote{The model was estimated on a time-period when the stock-bond correlation was broadly positive, hence the direction of the response. More recent data tends to show opposite correlation.}

\begin{figure}[!ht]
\centering
\includegraphics[width=1\textwidth]{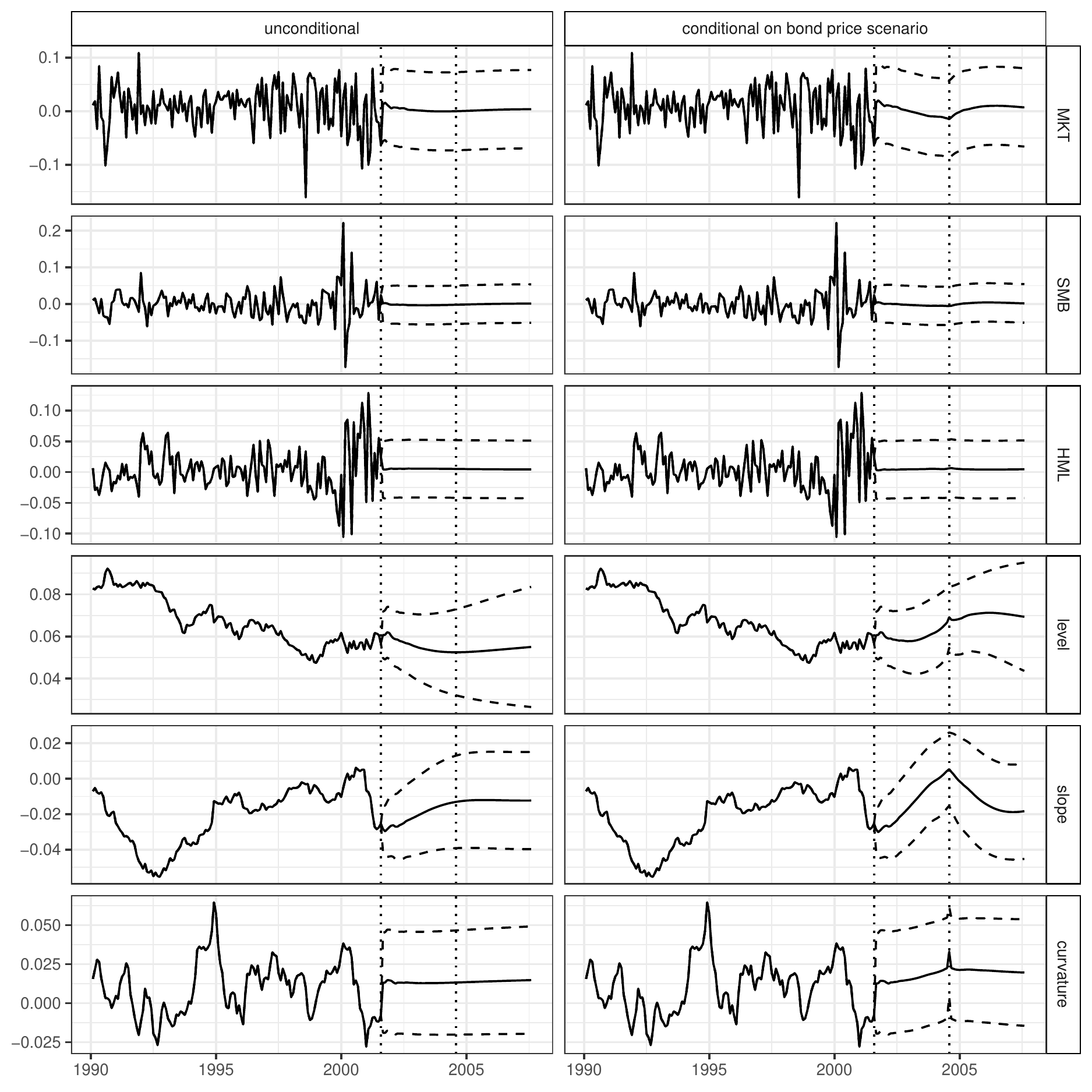}
\caption{Unconditional versus conditional (columns) predictions of the factor-level predictions (rows). The thick line plots the historical value up to August 2001 (first vertical dotted line) and the mean forecast thereafter. The dashed lines indicate a 90\% confidence interval. The timing of the conditional view is indicated by the second vertical dotted line.}
\label{fig:exampleBfactor}
\end{figure}

Figure \ref{fig:exampleBfactor} shows how these asset-level moves are explained by factor movement. Intuitively for the bond price to drop, the level, slope and curvature factor may all show an increase. We see that the change is mainly driven by the slope, which is intuitive since the level is generally more stable as it drives movements both at the long and the short end of the curve.

\begin{figure}[!ht]
\centering
\includegraphics[width=1\textwidth]{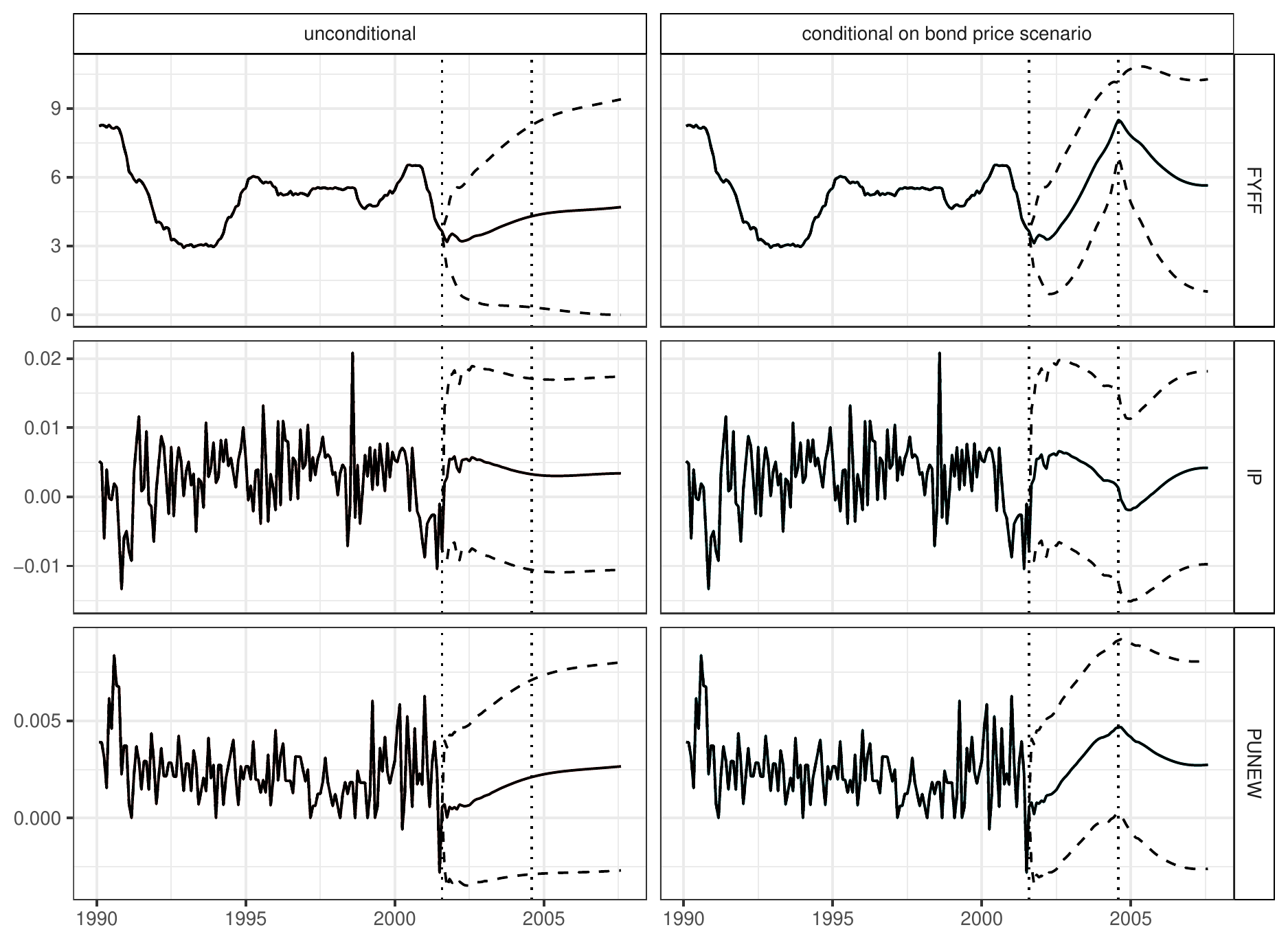}
\caption{Unconditional versus conditional (columns) predictions of the macro-level predictions (rows). The thick line plots the historical value up to August 2001 (first vertical dotted line) and the mean forecast thereafter. The dashed lines indicate a 90\% confidence interval. The timing of the conditional view is indicated by the second vertical dotted line.}
\label{fig:exampleBmacro}
\end{figure}

Figure \ref{fig:exampleBmacro} plots select macroeconomic variables. We can see the federal funds rate (FYFF) hike to explain the bond price movement, and in conjunction the industrial production (IP) drops relative to the baseline. Inflation measured in CPI (PUNEW) increases steadily at first, and then drops as rates come down. These movements are consistent with a cost-push shock \citep{steinsson2003optimal}.\footnote{For example, compare Figure \ref{fig:exampleBmacro} with \citet[second column in Figure 1]{ireland2011new}.}. The reverse stress test has thus identified that the unhedged liability is exposed to a cost-push macroeconomic scenario.

\section{Example B: DSGE with Nelson-Siegel factors}\label{sec:egC}

The second example uses the DSGE model of \citet{ireland2011new}, who analyses the latest three recessions of 1990, 2001 and 2008 from a New Keynesian perspective. Three variables, output, inflation and the nominal short rate are at the center of the analysis. DSGE models that are used in practice are oftentimes much larger, but with the goal of illustration in mind a more parsimonious model is suitable. The factor model reuses the Nelson-Siegel estimation from Section \ref{sec:egB}.

\subsection{Methodology}

This section discusses the methodology behind the macroeconomic model, the financial markets model, and the macro-financial link. I use the estimation approach explained in Section \ref{sub:estmacrofirst}, i.e.~to estimate the macroeconomic model first, under the assumption that the financial factors provide no additional information. The observations in the model measure the underlying shocks with no error, hence the model is identified and the assumption is valid. Instead of the original maximum likelihood estimation by \citet{ireland2011new}, I consider a Bayesian strategy to illustrate some additional features of the framework.

\subsubsection{Macroeconomic model}

For completeness, this section reiterates some of the results from \citet{ireland2011new}. The macroeconomic model is captured in seven (log-linearized) state equations,
\begin{align*}
(z-\beta\gamma)(z-\gamma)\hat{\lambda}_t&=\gamma z\hat{y}_{t-1}-(z^2-\beta\gamma^2)\hat{y}_t+\beta\gamma z\mathbb{E}_t[\hat{y}_{t+1}]+(z-\beta\gamma\rho_a)(z-\gamma)\hat{a}_t-\gamma z\hat{z}_t,\\
\hat{\lambda}_t&=\hat{r}_t+\mathbb{E}_t[\hat{\lambda}_{t+1}]-\mathbb{E}_t[\hat{\pi}_{t+1}],\\
(1+\beta\alpha)\hat{\pi}_t&=\alpha\hat{\pi}_{t-1}+\beta\mathbb{E}_t[\hat{\pi}_{t+1}]-\psi\hat{\lambda}_t+\psi\hat{a}_t+\hat{e}_t,\\
\hat{g}_t&=\hat{y}_t-\hat{y}_{t-1}+\hat{z}_t,\\
0&=\gamma z\hat{q}_{t-1}-(z^2+\beta\gamma^2)\hat{q}_t+\beta\gamma z\mathbb{E}_t[\hat{q}_{t+1}]+\beta\gamma(z-\gamma)(1-\rho_a)\hat{a}_t-\gamma z\hat{z}_t,\\
\hat{x}_t&=\hat{y}_t-\hat{q}_t,\\
\hat{r}_t&=\rho_r\hat{r}_{t-1}+\rho_\pi\hat{\pi}_t+\rho_g\hat{g}_t+\rho_x\hat{x}_t+\sigma_r\varepsilon_t^r.
\end{align*}
It also includes the following shocks
\begin{align*}
\hat{a}_t&=\rho_a\hat{a}_{t-1}+\sigma_a\varepsilon_t^a,\\
\hat{e}_t&=\rho_e\hat{e}_{t-1}+\sigma_e\varepsilon_t^e,\\
\hat{z}_t&=\sigma_z\varepsilon_t^z.
\end{align*}
The model is measured through three time series,
\begin{align*}
\hat{g}_t&=\ln(Y_t)-\ln(Y_t-1)-\ln(g),\\
\hat{\pi}_t&=\ln(P_t)-\ln(P_t-1)-\ln(\pi),\\
\hat{r}_t&=\ln(r_t)-\ln(r).
\end{align*}

I use the Bayesian estimation strategy outlined in Section \ref{sub:estmacrofirst}, i.e. to estimate the DSGE separately and in advance. The estimation procedure for the macroeconomic model on its own is an adaptive MCMC, with a chain length of $10^6$ after a burn-in of $10^5$. Every 100th draw is saved, so we have a sample of size $10^4$ from the posterior distribution. This procedure requires a prior distribution on the parameters of the model, and a way to compute the likelihood of observing the data given specific parameters. I assume the following relatively flat set of priors. The parameters $\alpha$, $\gamma$, $\rho_a$, $\rho_e$, $\rho_g$, $\rho_\pi$ are a priori uniformly distributed on the interval $[0,1]$, $\sigma_a$ is inverse-gamma distributed with mean $0.1$ and variance $1$, and $\sigma_e$, $\sigma_r$ and $\sigma_z$ are inverse-gamma distributed with mean $0.01$ and variance $1$.

The likelihood is computed as follows. For a specific set of parameters, and with the equations as specified above, I use the algorithm by \citet{anderson1985linear} to solve the system. The solved system can be written in DLM format of (\ref{eqn:dlm1}-\ref{eqn:dlm2}), which allows in turn for the log-likelihood computation using a standard Kalman filter.

\subsubsection{Financial markets model and macro-financial link}

As mentioned above, the factor model is the estimated Nelson-Siegel model outlined in Section \ref{sec:egBmethodology}. For the macro-financial link, I use the example in Section \ref{sub:estmacrofirst}, i.e. a flat normal-inverse-Wishart conjugate prior for the parameters in $\theta_f$.

\subsection{Data}

The data used in the original model is available on the web-appendix to the paper. It covers the real GDP, the GDP implicit price deflator, the 3-month treasury rate, and the US civilian population over age 16 (for normalization of the GDP) from 1983 to 2009. The data is complemented with the level, slope and curvature estimates from Section \ref{sec:egB}, converted to a quarterly frequency by taking the last month of each quarter, and extended in history to 2009 using updated treasury rate series.

\subsection{Results}

\begin{figure}[!ht]
\centering
\includegraphics[width=1\textwidth]{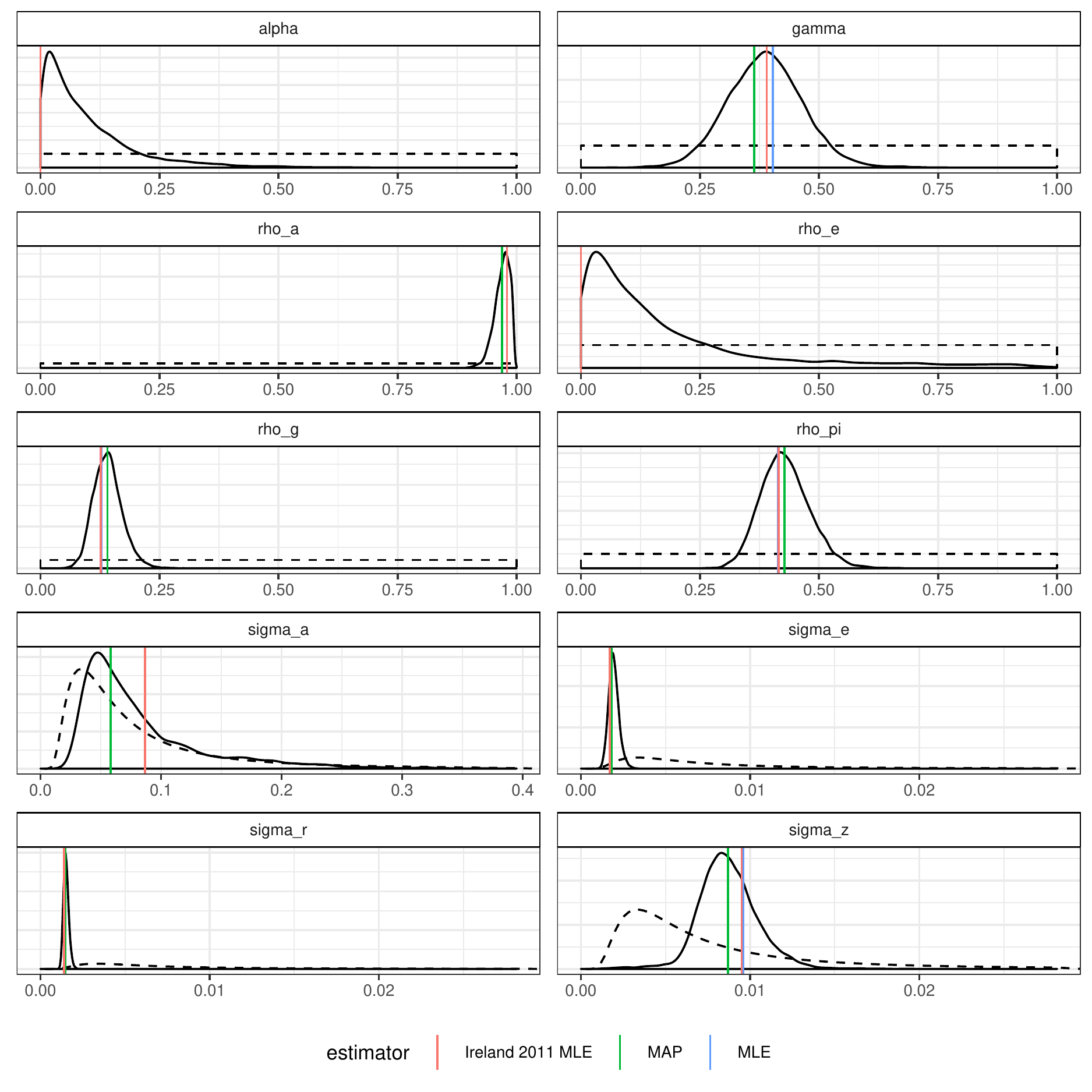}
\caption{MCMC estimates of the DSGE model parameter distributions (black), compared to priors (dashed), MAP estimate (red), and orginal MLE estimates in \citet[blue]{ireland2011new}.}
\label{fig:example_C_est}
\end{figure}

Figure \ref{fig:example_C_est} shows the marginal posterior distribution of the macro model parameters against their priors. It also includes the MLE estimates in the paper and the MAP estimate using a particle swarm optimizer. It is clear that the estimate by \citet{ireland2011new} is practically equivalent to the MLE estimate.\footnote{The small difference in the estimate of $\gamma$ likely stems from a difference in the implementation of the solver and the Kalman filter, or its starting point.}

I consider a scenario on the observed series as defined in the macroeconomic model. This is a joint scenario on two variables that represents a recession at a 5-year horizon. The recession itself is characterized by a -2\% quarterly GDP growth. Recessions may be demand or supply-shock driven. The demand-shock driven variant is identified through an additional view of 0\% on inflation in the same quarter.

\begin{figure}[!ht]
\centering
\includegraphics[width=1\textwidth]{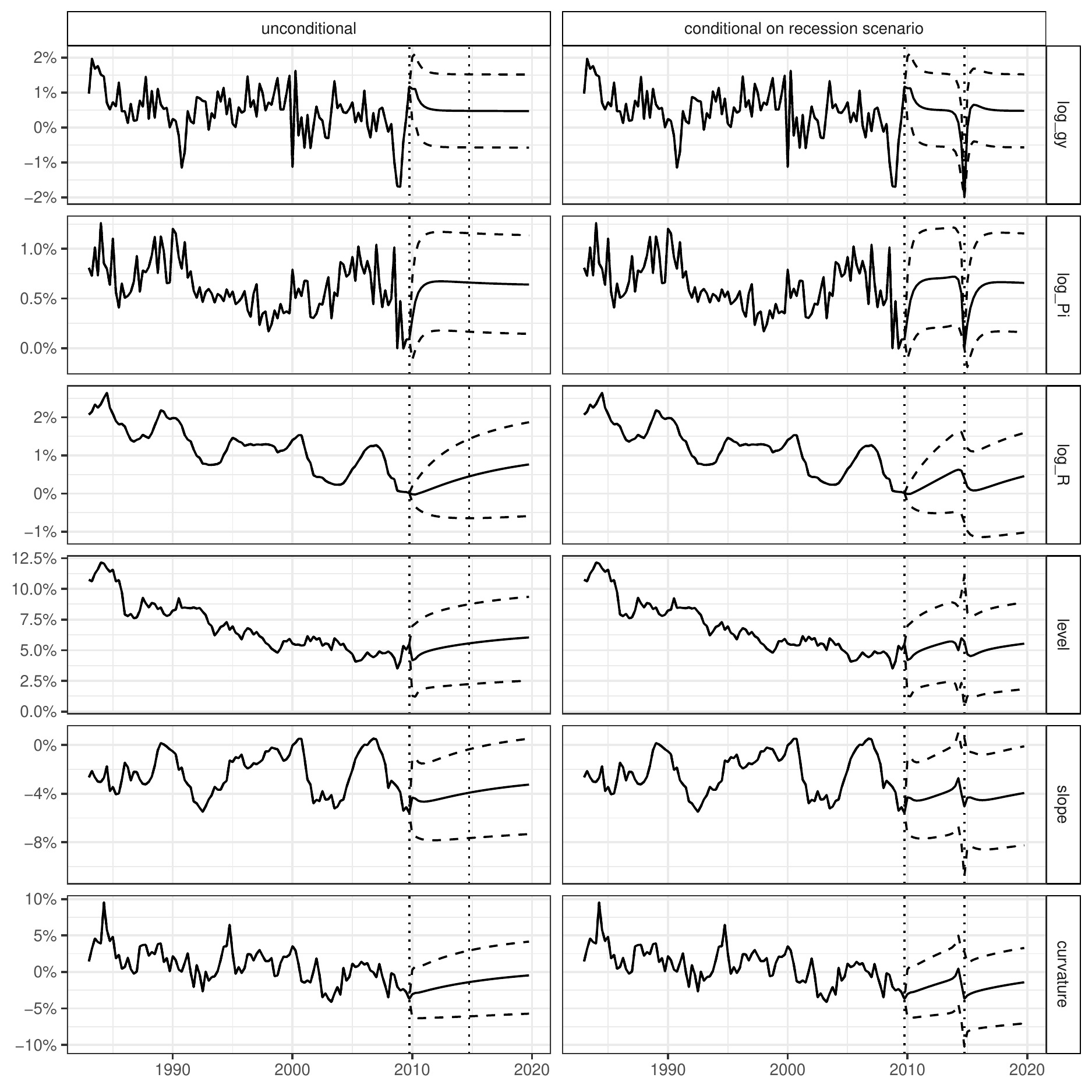}
\caption{Unconditional versus conditional (columns) predictions of the macro and factor-level predictions (rows). The thick line plots the historical value up to Q4 2009 (first vertical dotted line) and the mean forecast thereafter. The dashed lines indicate a 90\% confidence interval. The timing of the conditional views is indicated by the second vertical dotted line, and as the views are exact (no uncertainty), the confidence interval width of the corresponding variables shrinks to zero.}
\label{fig:example_C_measured_and_factor}
\end{figure}

Figure \ref{fig:example_C_measured_and_factor} shows the unconditional evolution of the macroeconomic observations and factor returns in the left column, versus the conditional case in the right column. We see a strong decline of interest rates as the FED tries to navigate the recession, and a recovery afterwards. The decline is visible in both rates as well as the level, slope and curvature factors, which all show GFC-like patterns. GDP growth is stronger than the baseline forecast right after the 5-year horizon, suggesting a recovery from the recessionary shock.

\section{Conclusion}\label{sec:conclusion}

Economic scenario generators should not be calibrated to historical data alone. For various purposes they need to be calibrated to stressed scenarios or expert views, or any other forward-looking information. There is no generally accepted way to do this. In finance there are methods available that allow for calibration of single-period variables, but these models are unfit for the multi-period macroeconomic scenarios that regulators prescribe. In macroeconomics, the approaches do not include enough granularity in financial variables to capture the level of detail that financial practitioners need. This paper proposes a conditional scenario simulation framework that marries the macroeconomic and the finance approach. Under certain econometric assumptions, the framework has a bring-your-own flexibility to macroeconomic and factor models. Two examples demonstrate how this would work for an empirical as well as a more theoretical macroeconomic model. Finally, for specific model choices and a single-period horizon, forecasting mean returns becomes equivalent to the Black-Litterman formula.

\appendix

\section{Forecasting procedure}\label{sec:forecasting}

This appendix describes the forecasting algorithm, conditional on the views. Future views are collected in the tuple $\psi$, and the dynamics of future assets in $\phi$. I distinguish two important cases for estimated parameters $\theta$ that control the macroeconomic model, the factor model and the macro-financial link. First a single estimate of $\theta$ may be available, as in Section \ref{sec:egB}. Second, we may have a sample of size $N$ available, $\theta^{(i)}$, $i=1,\ldots,N$, as in Section \ref{sec:egC}. I discuss both cases.

\subsection{Single estimate of $\theta$}\label{sub:singletheta}

All variables of interest can be expressed as linear combinations of $\tilde{x}_t$, $\alpha_t$ and $\varepsilon_t$, and it is tempting to collect all variables in a single DLM, along with all macroeconomic measurements, factor returns and views. However, $\alpha_t$ and $\varepsilon_t'''$ as well as views only exist in future states, whereas macroeconomic measurement and factor returns are only relevant in the past. The independence assumptions allow us to strictly split such a joint DLM up into a historical part running from $t=1,\ldots,T$, and a future part for $t=T\ldots,T+H$.\footnote{The overlapping period is not problematic, since conditioning can be applied in sequential processing steps on the same period for independent views \citep{koopman2000}.} 

The DLM for past data is as in (\ref{eqn:dlm3}-\ref{eqn:dlm4}). The second DLM combines the parameters in $\theta$, $\psi$ and $\phi$ into
\begin{align}\label{eqn:dlmfuture}
    \begin{bmatrix}\tilde{x}_{t+1}\\\alpha_{t+1}\end{bmatrix}
    &=\begin{bmatrix}A&O\\O&\Phi\end{bmatrix}
    \begin{bmatrix}\tilde{x}_t\\\alpha_t\end{bmatrix}
    +\begin{bmatrix}G&O&O&O\\O&O&S_t&O\end{bmatrix}
    \begin{bmatrix}\varepsilon'_t\\\varepsilon''_t\\\varepsilon'''_t\\\varepsilon''''_t\end{bmatrix},
    &\begin{bmatrix}\tilde{x}_T\\\alpha_T\end{bmatrix}&
    \sim\mathcal{N}\left(\begin{bmatrix}\tilde{x}_{T|T}\\0\end{bmatrix},
    \begin{bmatrix}P_{T|T}&O\\O&\tau\Sigma_T\end{bmatrix}\right),\nonumber
    \\\tilde{v}_t&=\begin{bmatrix}Q_t&&P_t\end{bmatrix}
    \begin{bmatrix}\tilde{x}_t\\\alpha_t\end{bmatrix}
    +\begin{bmatrix}R_t&\Omega_t^{\tfrac12}\end{bmatrix}
    \begin{bmatrix}\varepsilon_t\\\varepsilon''''_t\end{bmatrix},
    &\begin{bmatrix}\varepsilon_t\\\varepsilon''''_t\end{bmatrix}&\sim\mathcal{N}(0,I),
\end{align}
where $\Omega_t^{\tfrac12}\varepsilon_t''''\sim\xi_t\sim\mathcal{N}(0,\Omega_t)$ through LDL decomposition.\footnote{The dimension of $\varepsilon_t''''$ needs to be the largest rank of $\Omega_t$ of all times $t=T,\ldots,T+H$. $\Omega_t^{\tfrac12}$ can be column-appended with zeros to ensure the multiplication with $\varepsilon_t''''$ has the right dimension for all other times. In case there are no views at a given time, $\Omega_t$ is a $0\times0$ matrix and $\Omega_t^{\tfrac12}$ has zero rows.} Note how the initial distribution $(\tilde{x}_T,\alpha_T)$ can be determined from the final filtered value of the first DLM. That is, by applying the Kalman filter algorithm outlined by \citet{dejong1995}, which takes into account the correlation between state and measurement errors, we get estimates $\tilde{x}_{T|T}$ and $P_{T|T}$. 

Subsequently, apply the Kalman filter to the second DLM in (\ref{eqn:dlmfuture}) to get the filtered mean and covariance forecasts of $\tilde{x}_t$ and $\alpha_t$ jointly for all $t=T,\ldots,T+H$. These forecasts only include views up to the same time.

Finally, running the Kalman smoother backwards in time from $t=T+H,\ldots,T$ gives the updated means and joint covariances for
\begin{align*}
x_t,\alpha_t,\varepsilon_t&|y_{1:T},f_{1:T},v_{T:T+H},&t&=T,\ldots,T+H.
\end{align*}
Obtaining the variance of any linear combination of these variables is straightforward. Alternatively, we can simulate any number of paths from this distribution using the simulation smoother, which is also described in \citet{dejong1995}.

\subsection{Sample from the distribution $\theta$}

In case we have an entire sample of size $N$ available, $\theta^{(i)}$, $i=1,\ldots,N$, it is possible to use the algorithms above. If the interest is in simulation, simply construct the DLMs above for each parameterization $\theta^{(i)}$, and simulate a single path.

Naturally, this approach can also be used to approximate the analytical distribution. However, it is more efficient to rely on the laws of total expectation and variance. Suppose that through filtering and smoothing, vectors $\hat{c}_{t|T+H}(\theta)$ and $\sigma_{t|T+H}^2(\theta)$ are obtained, containing the means and variances of a vector of all variables of interest, $c_t$ for $t=T,\ldots,T+H$, for a specific value $\theta$. Then, with expectations and variances conditional on all data, $y_{1:T}$, $f_{1:T}$, and $v_{T:T+H}$,
\begin{align*}
\E[c_t]&=\E[\E[c_t|\theta]]=\E[\hat{c}_{t|T+H}(\theta)],\\
\mathrm{Var}(c_t)&=\E[\mathrm{Var}(c_t|\theta)]+\mathrm{Var}(\E[c_t|\theta])
=\E[\sigma_{t|T+H}^2(\theta)]+\mathrm{Var}(\hat{c}_{t|T+H}(\theta)).
\end{align*}
Finally, take the sample mean and sample variance over all $\theta^{(i)}$ to approximate the mean and variance.

\section{Link to Black-Litterman}\label{sec:blacklitterman}

The CSG and the BL model overlap as both prescribe an unconditional distribution of mean returns, which is then updated with views. To show that the BL model is a special case of the CSG, I construct the CSG from a specific macro and factor model, which is then analysed for a single period to show that the forecasts align. This appendix uses the notation introduced in Section \ref{sec:notation}.

I start with the introduction of the BL model. For ease of comparison, we use a slightly different formulation. \citet{black1992} do not include the risk-free rate in their analysis, and instead consider all returns to be excess returns. In our comparison we will add a constant risk-free rate $r_f$ to all returns.\footnote{I do this since the entire notation of this paper has defined means as non-excess. The analysis in this appendix works as well for the excess return case, by simply substituting $r_f=0$ in what follows.} The prior distribution of mean returns (given no views) is defined as
\begin{align}
    \mu\sim\mathcal{N}(r_f+\Sigma w\lambda,\tau\Sigma),
\end{align}
where $\Sigma$ is the covariance matrix of all available assets, $w$ are the weights of the market portfolio, $\lambda$ is the market price of risk, and $\tau$ is a scaling parameter that defined the tightness of the prior.

With views $v_t$ on the returns of the following format
\begin{align}
v&=P\mu+\xi,&\xi\sim\mathcal{N}(0,\Omega),
\end{align}
after applying the BL formula, the updated returns are
\begin{align}\label{eqn:blacklitterman}
\mu|v&\sim\mathcal{N}(((\tau\Sigma)^{-1}+P^\top\Omega^{-1}P)^{-1}((\tau\Sigma)^{-1}\pi+P^\top\Omega^{-1}v),((\tau\Sigma)^{-1}+P^\top\Omega^{-1}P)^{-1}),
\end{align}
where $\pi=r_f+\Sigma w\lambda$ is the prior mean. 

I now show how this formula is a special case of the CSG in this paper, by setting up a specific implementation with views, that results in the exact same posterior distribution of the asset mean returns. Set $t=T=1$ and $H=0$, i.e.~we only analyse the CSG for a single period, as the BL model is not multi-period. 

The macroeconomic model has a single variable, namely the risk aversion parameter $\lambda_t$, which remains constant over time and equal to $\lambda$. The factor model has two factors, a (known and constant) risk-free rate $r_t^f=r_f$ and a market factor $m_t=w^\top(r_t-r_t^f)$ in excess of the risk-free rate, with mean $\mu_t^m$ and variance $\sigma_m^2$. Suppose all assets in the market are included, then with $w$ the weights of the market portfolio, from CAPM it follows that
\[\mu_t=\alpha_t+r_t^f+\frac{\mathrm{cov}(r_t, r_t^m)}{\sigma_m^2}\mu^m_t
=\alpha_t+\underbrace{\begin{bmatrix}1&\Sigma w\sigma_m^{-2}\end{bmatrix}}_{\beta_t}
\begin{bmatrix}r_t^f\\\mu_t^m\end{bmatrix}.\]
From standard regression results we also get that
\begin{align*}
    \Sigma^r
    &=\mathrm{vcov}(r_t-\beta f)=\mathrm{vcov}(r_t-r_t^f-\Sigma w\sigma^{-2}_mw^\top(r_t-r_t^f))
    =\mathrm{vcov}((I-\Sigma w\sigma^{-2}_mw^\top)r_t)\\
    &=\Sigma-2\sigma^{-2}_m\Sigma ww^\top\Sigma+\sigma^{-4}_m\Sigma ww^\top\Sigma ww^\top\Sigma
    =\Sigma-\sigma^{-2}_m\Sigma ww^\top\Sigma.
\end{align*}
As a consequence, by choosing the initial distribution of the macroeconomic variable as $\lambda_t\sim\mathcal{N}(\lambda,\tau\sigma_m^{-2})$ and the macro-financial linkage as $\mu_t^m=\sigma_m^2\lambda_t$, the BL unconditional (prior) distribution follows as defined by joint normality and
\begin{align*}
    \hat{\mu}_{1|0}&=\E[\mu_t]=\E[\alpha_1]+r_1^f+\Sigma w\sigma_m^{-2}\E[\sigma_m^2\lambda_1]
    =0+r_f+\Sigma w\lambda
    =\pi,\\
    \Sigma_{1|0}&=\mathrm{vcov}(\mu_t)=\mathrm{vcov}(\alpha_1)+\beta\mathrm{vcov}(\mu_1^f)\beta^\top
    =\tau\Sigma^r+\Sigma w\sigma_m^{-2}\mathrm{var}(\mu_t^m)\sigma_m^{-2}w^\top\Sigma
    =\tau\Sigma.
\end{align*}

Note that implicitly, all parameters in the tuple $\theta$ are defined as
\begin{align*}
    A&=1,&B&=1,&\Gamma&=\begin{bmatrix}0\\\sigma^2_w\end{bmatrix},&G&=0,&H&=0,&F&=\begin{bmatrix}0\\\sigma_w\end{bmatrix},&\bar{x}&=0,&\bar{y}&=0,&\bar{f}&=\begin{bmatrix}r_f\\0\end{bmatrix}.
\end{align*}
Section \ref{sub:conditional} shows how to define the parameter in the tuple $\phi$ for views on asset mean returns. We start from $v_t^\mu=P_t^\mu\mu_t+\xi_t^\mu$, where $P_t^\mu=P$. For such views $\bar{v}_t=P_t^\mu\beta_t\bar{f}=Pr_f$ and $\phi$ is defined through
\begin{align*}
    P_t&=P_t^\mu=P,&Q_t&=P_t^\mu\beta_t\Gamma=P\Sigma w,&R_t&=0,&\Omega_t&=\Omega.
\end{align*}

To prove equivalence after updating the mean returns with the views, we need to use compute the forecasts in accordance with Appendix \ref{sub:singletheta}. Since we know the distribution of the macroeconomic variable at $t=T=1$, i.e.~$\lambda_t\sim\mathcal{N}(\lambda,\tau\sigma^{-2}_m)$, we only need a single Kalman filter update step to get the posterior distribution $\mu_1|v_1$.\footnote{An additional smoothing stage is not necessary, as the last filtering step already includes all available future data. Smoothing does not affect the last time step.} Running through a single step of the Kalman filter gives (with $S_t$ the innovation covariance, and $K_t$ the optimal Kalman gain)
\begin{align*}
    \tilde{y}_1&=v_1-\bar{v}_1-\begin{bmatrix}Q_1&P_1\end{bmatrix}\begin{bmatrix}\lambda\\0\end{bmatrix}
    =v-P(r_f+\Sigma w\lambda)
    =v-P\pi\\
    S_1&=\begin{bmatrix}Q_1&P_1\end{bmatrix}\begin{bmatrix}\tau\sigma_m^{-2}&0^\top\\0&\tau\Sigma^r\end{bmatrix}\begin{bmatrix}Q_1^\top\\P_1^\top\end{bmatrix}+\Omega_1
    =P\Sigma w\tau\sigma_m^{-2}w^\top\Sigma P^\top+P\tau\Sigma^rP^\top+\Omega
    \\&=P\tau(\Sigma-\Sigma^r)P^\top+P\tau\Sigma^rP^\top+\Omega
    =P\tau\Sigma P^\top+\Omega\\
    K_1&=\begin{bmatrix}\tau\sigma_m^{-2}&0^\top\\0&\tau\Sigma^r\end{bmatrix}\begin{bmatrix}Q_1^\top\\P_1^\top\end{bmatrix}S_1^{-1}
    =\tau\begin{bmatrix}\sigma_m^{-2}w^\top\Sigma\\\Sigma^r\end{bmatrix}P^\top(P\tau\Sigma P^\top+\Omega)^{-1}\\
    \tilde{x}_{1|1}&=\begin{bmatrix}\lambda\\0\end{bmatrix}+K_1\tilde{y}_1
    =\begin{bmatrix}\lambda\\0\end{bmatrix}+\tau\begin{bmatrix}\sigma_m^{-2}w^\top\Sigma\\\Sigma^r\end{bmatrix}P^\top(P\tau\Sigma P^\top+\Omega)^{-1}(v-P\pi)\\
    P_{1|1}&=(I-K_1\begin{bmatrix}Q_1&P_1\end{bmatrix})\begin{bmatrix}\tau\sigma_m^{-2}&0^\top\\0&\tau\Sigma^r\end{bmatrix}
    \\&=\tau\begin{bmatrix}\sigma_m^{-2}&0^\top\\0&\Sigma^r\end{bmatrix}-\tau\begin{bmatrix}\sigma_m^{-2}w^\top\Sigma\\\Sigma^r\end{bmatrix}P^\top(P\tau\Sigma P^\top+\Omega)^{-1}P\tau\begin{bmatrix}\sigma_m^{-2}\Sigma w&\Sigma^r\end{bmatrix}
\end{align*}
Multiplying with $\begin{bmatrix}\beta_1\Gamma&I\end{bmatrix}=\begin{bmatrix}\Sigma w&I\end{bmatrix}$ and adding factor means gives the posterior mean and covariance matrix of the mean returns.
\begin{align*}
    \mu_{1|1}&=\beta_1\bar{f}+\begin{bmatrix}\beta_1\Gamma&I\end{bmatrix}\tilde{x}_{1|1}
    \\&=r_f+\beta\Gamma\lambda+\tau(\Sigma w\sigma_m^{-2}w^\top\Sigma+\Sigma^r)P^\top(P\tau\Sigma P^\top+\Omega)^{-1}(v-P\pi))
    \\&=r_f+\Sigma w\lambda+\tau\Sigma P^\top(P\tau\Sigma P^\top+\Omega)^{-1}(v-P\pi),\\
    \Sigma_{1|1}&=\begin{bmatrix}\Sigma w&I\end{bmatrix}P_{1|1}\begin{bmatrix}w^\top\Sigma\\I\end{bmatrix}
    =\tau\Sigma-\tau\Sigma P^\top(P\tau\Sigma P^\top+\Omega)^{-1}P\tau\Sigma.
\end{align*}
Although this posterior mean and covariance matrix look different from the BL formula in (\ref{eqn:blacklitterman}), they are in fact identical. This can be seen by applying the Woodbury matrix identity on the posterior mean and covariance matrix in (\ref{eqn:blacklitterman}). Starting with the covariance,
\begin{align*}
    ((\tau\Sigma)^{-1}+P^\top\Omega^{-1}P)^{-1})&=
    \tau\Sigma-\tau\Sigma P^\top(\Omega+P\tau\Sigma P^\top)^{-1}P\tau\Sigma=\Sigma_{1|1},
\end{align*}
and reusing this result for the mean shows that also
\begin{align*}
    &((\tau\Sigma)^{-1}+P^\top\Omega^{-1}P)^{-1}((\tau\Sigma)^{-1}\pi+P^\top\Omega^{-1}v)
    \\&\quad=(\tau\Sigma-\tau\Sigma P^\top(\Omega+P\tau\Sigma P^\top)^{-1}P\tau\Sigma)((\tau\Sigma)^{-1}\pi+P^\top\Omega^{-1}v)
    \\&\quad=\pi
    -\tau\Sigma P^\top(\Omega+P\tau\Sigma P^\top)^{-1}P\pi
    +\tau\Sigma P^\top\Omega^{-1}v
    -\tau\Sigma P^\top(\Omega+P\tau\Sigma P^\top)^{-1}P\tau\Sigma P^\top\Omega^{-1}v
    \\&\quad=\pi
    -\tau\Sigma P^\top(\Omega+P\tau\Sigma P^\top)^{-1}P\pi
    +\tau\Sigma P^\top(\Omega+P\tau\Sigma P^\top)^{-1}
    (\Omega+P\tau\Sigma P^\top-P\tau\Sigma P^\top)\Omega^{-1}v
    \\&\quad=\pi
    -\tau\Sigma P^\top(\Omega+P\tau\Sigma P^\top)^{-1}(P\pi-v)
    =\mu_{1|1},
\end{align*}
which completes the proof of equivalence.

\bibliography{literature}

\begin{thebibliography}{}

\bibitem[Acharya et~al., 2012]{acharya2012capital}
Acharya, V., Engle, R., and Richardson, M. (2012).
\newblock Capital shortfall: A new approach to ranking and regulating systemic
  risks.
\newblock {\em American Economic Review}, 102(3):59--64.

\bibitem[Almgren and Chriss, 2007]{almgren2007optimal}
Almgren, R. and Chriss, N. (2007).
\newblock Optimal portfolios from ordering information.
\newblock In {\em Forecasting Expected Returns in the Financial Markets}, pages
  55--100. Elsevier.

\bibitem[Anderson and Moore, 1985]{anderson1985linear}
Anderson, G. and Moore, G. (1985).
\newblock A linear algebraic procedure for solving linear perfect foresight
  models.
\newblock {\em Economics letters}, 17(3):247--252.

\bibitem[Ang and Bekaert, 2006]{ang2006}
Ang, A. and Bekaert, G. (2006).
\newblock Stock return predictability: Is it there?
\newblock {\em The Review of Financial Studies}, 20(3):651--707.

\bibitem[Ang and Piazzesi, 2003]{ang2003}
Ang, A. and Piazzesi, M. (2003).
\newblock A no-arbitrage vector autoregression of term structure dynamics with
  macroeconomic and latent variables.
\newblock {\em Journal of Monetary economics}, 50(4):745--787.

\bibitem[Arias et~al., 2014]{arias2014inference}
Arias, J., Rubio-Ramirez, J.~F., and Waggoner, D.~F. (2014).
\newblock Inference based on {SVAR} identified with sign and zero restrictions:
  Theory and applications.
\newblock CEPR Discussion Paper Series No. DP9796, Centre for Economic Policy
  Research.

\bibitem[Ba{\'n}bura et~al., 2015]{banbura2015conditional}
Ba{\'n}bura, M., Giannone, D., and Lenza, M. (2015).
\newblock Conditional forecasts and scenario analysis with vector
  autoregressions for large cross-sections.
\newblock {\em International Journal of forecasting}, 31(3):739--756.

\bibitem[Barras et~al., 2010]{barras2010false}
Barras, L., Scaillet, O., and Wermers, R. (2010).
\newblock False discoveries in mutual fund performance: Measuring luck in
  estimated alphas.
\newblock {\em The journal of finance}, 65(1):179--216.

\bibitem[Bernanke et~al., 2005]{bernanke2005measuring}
Bernanke, B.~S., Boivin, J., and Eliasz, P. (2005).
\newblock Measuring the effects of monetary policy: A factor-augmented vector
  autoregressive ({FAVAR}) approach.
\newblock {\em The Quarterly journal of economics}, 120(1):387--422.

\bibitem[Black and Litterman, 1992]{black1992}
Black, F. and Litterman, R. (1992).
\newblock Global portfolio optimization.
\newblock {\em Financial analysts journal}, 48(5):28--43.

\bibitem[Blanchard and Quah, 1989]{blanchard1989dynamic}
Blanchard, O.~J. and Quah, D. (1989).
\newblock The dynamic effects of aggregate demand and supply disturbances.
\newblock {\em The American Economic Review}, 79(4):655--673.

\bibitem[Bloor and Matheson, 2011]{bloor2011real}
Bloor, C. and Matheson, T. (2011).
\newblock Real-time conditional forecasts with {Bayesian VARs}: An application
  to new zealand.
\newblock {\em The North American Journal of Economics and Finance},
  22(1):26--42.

\bibitem[Boivin and Giannoni, 2006]{boivin2006dsge}
Boivin, J. and Giannoni, M. (2006).
\newblock {DSGE} models in a data-rich environment.
\newblock NBER Working Paper Series No. 12772, National Bureau of Economic
  Research.

\bibitem[Breuer et~al., 2012]{breuer2012systematic}
Breuer, T., Janda{\v{c}}ka, M., Menc{\'\i}a, J., and Summer, M. (2012).
\newblock A systematic approach to multi-period stress testing of portfolio
  credit risk.
\newblock {\em Journal of Banking \& Finance}, 36(2):332--340.

\bibitem[Busse et~al., 2010]{busse2010performance}
Busse, J.~A., Goyal, A., and Wahal, S. (2010).
\newblock Performance and persistence in institutional investment management.
\newblock {\em The Journal of Finance}, 65(2):765--790.

\bibitem[Campbell, 1987]{campbell1987stock}
Campbell, J.~Y. (1987).
\newblock Stock returns and the term structure.
\newblock {\em Journal of financial economics}, 18(2):373--399.

\bibitem[Campbell, 2003]{campbell2003consumption}
Campbell, J.~Y. (2003).
\newblock Consumption-based asset pricing.
\newblock {\em Handbook of the Economics of Finance}, 1:803--887.

\bibitem[Campbell and Shiller, 1988a]{campbell1988b}
Campbell, J.~Y. and Shiller, R.~J. (1988a).
\newblock The dividend-price ratio and expectations of future dividends and
  discount factors.
\newblock {\em The Review of Financial Studies}, 1(3):195--228.

\bibitem[Campbell and Shiller, 1988b]{campbell1988a}
Campbell, J.~Y. and Shiller, R.~J. (1988b).
\newblock Stock prices, earnings, and expected dividends.
\newblock {\em The Journal of Finance}, 43(3):661--676.

\bibitem[Campbell and Vuolteenaho, 2004]{campbell2004inflation}
Campbell, J.~Y. and Vuolteenaho, T. (2004).
\newblock Inflation illusion and stock prices.
\newblock {\em American Economic Review}, 94(2):19--23.

\bibitem[Chen et~al., 1986]{chen1986economic}
Chen, N.-F., Roll, R., and Ross, S.~A. (1986).
\newblock Economic forces and the stock market.
\newblock {\em Journal of business}, pages 383--403.

\bibitem[Christiano et~al., 1999]{christiano1999monetary}
Christiano, L.~J., Eichenbaum, M., and Evans, C.~L. (1999).
\newblock Monetary policy shocks: What have we learned and to what end?
\newblock {\em Handbook of macroeconomics}, 1:65--148.

\bibitem[Claessens and Kose, 2018]{claessens2018}
Claessens, S. and Kose, M.~A. (2018).
\newblock Frontiers of macrofinancial linkages.

\bibitem[Clarida and Coyle, 1984]{clarida1984conditional}
Clarida, R.~H. and Coyle, D. (1984).
\newblock Conditional projection by means of {Kalman} filtering.
\newblock NBER Technical Paper Series No.~36, National Bureau of Economic
  Research.

\bibitem[Clemen and Winkler, 1999]{clemen1999combining}
Clemen, R.~T. and Winkler, R.~L. (1999).
\newblock Combining probability distributions from experts in risk analysis.
\newblock {\em Risk analysis}, 19(2):187--203.

\bibitem[Cochrane, 2011]{cochrane2011presidential}
Cochrane, J.~H. (2011).
\newblock Presidential address: Discount rates.
\newblock {\em The Journal of finance}, 66(4):1047--1108.

\bibitem[Cole and McCullough, 2014]{cole2014basis}
Cole, C. and McCullough, K. (2014).
\newblock Basis risk, procyclicality, and systemic risk in the {Solvency II}
  equity risk module.
\newblock {\em Journal of Insurance Regulation}, 33(1):1--39.

\bibitem[De~Jong and Shephard, 1995]{dejong1995}
De~Jong, P. and Shephard, N. (1995).
\newblock The simulation smoother for time series models.
\newblock {\em Biometrika}, 82(2):339--350.

\bibitem[Durbin and Koopman, 2002]{durbin2002simple}
Durbin, J. and Koopman, S.~J. (2002).
\newblock A simple and efficient simulation smoother for state space time
  series analysis.
\newblock {\em Biometrika}, 89(3):603--616.

\bibitem[Fama and French, 1989]{fama1989business}
Fama, E.~F. and French, K.~R. (1989).
\newblock Business conditions and expected returns on stocks and bonds.
\newblock {\em Journal of financial economics}, 25(1):23--49.

\bibitem[Fama and French, 1992]{fama1992}
Fama, E.~F. and French, K.~R. (1992).
\newblock The cross-section of expected stock returns.
\newblock {\em the Journal of Finance}, 47(2):427--465.

\bibitem[Fama and French, 1993]{fama1993}
Fama, E.~F. and French, K.~R. (1993).
\newblock Common risk factors in the returns on stocks and bonds.
\newblock {\em Journal of financial economics}, 33(1):3--56.

\bibitem[Fama and French, 2010]{fama2010luck}
Fama, E.~F. and French, K.~R. (2010).
\newblock Luck versus skill in the cross-section of mutual fund returns.
\newblock {\em The journal of finance}, 65(5):1915--1947.

\bibitem[Ferson and Schadt, 1996]{ferson1996measuring}
Ferson, W.~E. and Schadt, R.~W. (1996).
\newblock Measuring fund strategy and performance in changing economic
  conditions.
\newblock {\em The Journal of finance}, 51(2):425--461.

\bibitem[Fisher, 2006]{fisher2006dynamic}
Fisher, J.~D. (2006).
\newblock The dynamic effects of neutral and investment-specific technology
  shocks.
\newblock {\em Journal of political Economy}, 114(3):413--451.

\bibitem[French, 2019]{french2019data}
French, K.~R. (2019).
\newblock Data library.
\newblock data retrieved from Kenneth R. French - Library,
  \url{https://mba.tuck.dartmouth.edu/pages/faculty/ken.french/data_library.html}.

\bibitem[Giannone et~al., 2014]{giannone2014short}
Giannone, D., Lenza, M., Momferatou, D., and Onorante, L. (2014).
\newblock Short-term inflation projections: A {Bayesian} vector autoregressive
  approach.
\newblock {\em International journal of forecasting}, 30(3):635--644.

\bibitem[Giannone et~al., 2012]{giannone2012ecb}
Giannone, D., Lenza, M., Pill, H., and Reichlin, L. (2012).
\newblock The {ECB} and the interbank market.
\newblock {\em The Economic Journal}, 122(564):F467--F486.

\bibitem[Giannone et~al., 2010]{giannone2008business}
Giannone, D., Lenza, M., and Reichlin, L. (2010).
\newblock {\em Business Cycles in the Euro Area}, pages 141--167.
\newblock University of Chicago Press.

\bibitem[Golub et~al., 2018]{golub2018market}
Golub, B., Greenberg, D., and Ratcliffe, R. (2018).
\newblock Market-driven scenarios: An approach for plausible scenario
  construction.
\newblock {\em The Journal of Portfolio Management}, 44(5):6--20.

\bibitem[Grundke, 2011]{grundke2011reverse}
Grundke, P. (2011).
\newblock Reverse stress tests with bottom-up approaches.
\newblock {\em The Journal of Risk Model Validation}, 5(1):71--90.

\bibitem[Ha et~al., 2020]{ha2020global}
Ha, J., Kose, M.~A., Otrok, C., and Prasad, E.~S. (2020).
\newblock Global macro-financial cycles and spillovers.
\newblock NBER Working Paper Series No. 26798, National Bureau of Economic
  Research.

\bibitem[Ireland, 2011]{ireland2011new}
Ireland, P.~N. (2011).
\newblock A new {Keynesian} perspective on the great recession.
\newblock {\em Journal of Money, Credit and Banking}, 43(1):31--54.

\bibitem[Jarocinski and Smets, 2008]{jarocinski2008house}
Jarocinski, M. and Smets, F. (2008).
\newblock House prices and the stance of monetary policy.
\newblock ECB Working Paper Series No. 891, European Central Bank.

\bibitem[Karlsson, 2013]{karlsson2013forecasting}
Karlsson, S. (2013).
\newblock Forecasting with {Bayesian} vector autoregression.
\newblock In {\em Handbook of economic forecasting}, volume~2, pages 791--897.
  Elsevier.

\bibitem[Koopman and Durbin, 2000]{koopman2000}
Koopman, S.~J. and Durbin, J. (2000).
\newblock Fast filtering and smoothing for multivariate state space models.
\newblock {\em Journal of Time Series Analysis}, 21(3):281--296.

\bibitem[Kupiec, 2002]{kupiec2002stress}
Kupiec, P. (2002).
\newblock Stress testing in a value at risk framework.
\newblock {\em Risk Management: Value at Risk and Beyond, ed. by M. Dempster},
  pages 76--99.

\bibitem[Lenza et~al., 2010]{lenza2010monetary}
Lenza, M., Pill, H., and Reichlin, L. (2010).
\newblock Monetary policy in exceptional times.
\newblock {\em Economic Policy}, 25(62):295--339.

\bibitem[Mamaysky et~al., 2008]{mamaysky2008estimating}
Mamaysky, H., Spiegel, M., and Zhang, H. (2008).
\newblock Estimating the dynamics of mutual fund alphas and betas.
\newblock {\em The Review of Financial Studies}, 21(1):233--264.

\bibitem[Meucci, 2008]{meucci2008fully}
Meucci, A. (2008).
\newblock Fully flexible views: Theory and practice.
\newblock {\em Fully Flexible Views: Theory and Practice, Risk},
  21(10):97--102.

\bibitem[Meucci, 2009]{meucci2009}
Meucci, A. (2009).
\newblock Enhancing the {Black--Litterman} and related approaches: Views and
  stress-test on risk factors.
\newblock {\em Journal of Asset Management}, 10(2):89--96.

\bibitem[Meucci, 2010]{meucci2010}
Meucci, A. (2010).
\newblock The {Black--Litterman} approach.
\newblock {\em Encyclopedia of Quantitative Finance}.

\bibitem[Nelson and Siegel, 1987]{nelson1987}
Nelson, C.~R. and Siegel, A.~F. (1987).
\newblock Parsimonious modeling of yield curves.
\newblock {\em Journal of business}, pages 473--489.

\bibitem[Palczewski and Palczewski, 2019]{palczewski2019black}
Palczewski, A. and Palczewski, J. (2019).
\newblock {Black--Litterman} model for continuous distributions.
\newblock {\em European Journal of Operational Research}, 273(2):708--720.

\bibitem[Pezier, 2007]{pezier2007global}
Pezier, J. (2007).
\newblock Global portfolio optiomization revisted: A least discrimination
  alternative to {Black--Litterman}.
\newblock Technical report, Henley Business School, Reading University.

\bibitem[Qian and Gorman, 2001]{qian2001conditional}
Qian, E. and Gorman, S. (2001).
\newblock Conditional distribution in portfolio theory.
\newblock {\em Financial Analysts Journal}, 57(2):44--51.

\bibitem[Sheikh, 1996]{sheikh1996barra}
Sheikh, A. (1996).
\newblock Barra's risk models.
\newblock {\em Barra Research Insights}, pages 1--24.

\bibitem[Sims, 1980]{sims1980macroeconomics}
Sims, C.~A. (1980).
\newblock Macroeconomics and reality.
\newblock {\em Econometrica: journal of the Econometric Society}, pages 1--48.

\bibitem[Steinsson, 2003]{steinsson2003optimal}
Steinsson, J. (2003).
\newblock Optimal monetary policy in an economy with inflation persistence.
\newblock {\em Journal of Monetary Economics}, 50(7):1425--1456.

\bibitem[Uhlig, 2005]{uhlig2005effects}
Uhlig, H. (2005).
\newblock What are the effects of monetary policy on output? {R}esults from an
  agnostic identification procedure.
\newblock {\em Journal of Monetary Economics}, 52(2):381--419.

\bibitem[Van~der Schans and Steehouwer, 2017]{vanderschans2017}
Van~der Schans, M. and Steehouwer, H. (2017).
\newblock Time-dependent {Black--Litterman}.
\newblock {\em Journal of Asset Management}, 18(5):371--387.

\bibitem[Waggoner and Zha, 1999]{waggoner1999conditional}
Waggoner, D.~F. and Zha, T. (1999).
\newblock Conditional forecasts in dynamic multivariate models.
\newblock {\em Review of Economics and Statistics}, 81(4):639--651.

\bibitem[Welch and Goyal, 2007]{welch2007comprehensive}
Welch, I. and Goyal, A. (2007).
\newblock A comprehensive look at the empirical performance of equity premium
  prediction.
\newblock {\em The Review of Financial Studies}, 21(4):1455--1508.

\end{thebibliography}

\end{document}